\documentclass[a4paper,onecolumn,11pt,accepted=2023-05-11]{quantumarticle}
\pdfoutput=1
\usepackage[utf8]{inputenc}
\usepackage[english]{babel}
\usepackage[T1]{fontenc}
\usepackage{amsmath,amssymb,amsfonts,dsfont}
\usepackage{hyperref,centernot}

\usepackage[numbers,sort&compress]{natbib}

\usepackage{tikz}
\usepackage{lipsum}

\newcommand{\<}{\langle}
\renewcommand{\>}{\rangle}

\newcommand{\norm}[1]{\lVert#1\rVert}

\providecommand{\tr}{{\rm Tr}}
\renewcommand{\phi}{\varphi}
\newcommand{\bra}[1]{\langle #1\rvert}
\newcommand{\ket}[1]{\lvert #1\rangle}

\begin{document}

\title{On Quantum Steering and Wigner Negativity}

\author{Mattia Walschaers}
\email{mattia.walschaers@lkb.upmc.fr}
\affiliation{Laboratoire Kastler Brossel, Sorbonne Universit\'{e}, CNRS, ENS-Universit\'{e} PSL,  Coll\`{e}ge de France, 4 place Jussieu, F-75252 Paris, France}
\orcid{0000-0003-4649-0044}

\maketitle

\begin{abstract}

Quantum correlations and Wigner negativity are two important signatures of nonclassicality in continuous-variable quantum systems. In this work, we investigate how both are intertwined in the context of the conditional generation of Wigner negativity. It was previously shown that when Alice and Bob share a Gaussian state, Bob can perform some measurement on his system to create Wigner negativity on Alice's side if and only if there is Gaussian steering from Alice to Bob. In this work, we attempt to generalise these findings to a much broader class of scenarios on which Alice and Bob share a non-Gaussian state. We show that if Alice can initially steer Bob's system with Wigner-positive measurements, Bob can remotely create Wigner negativity in Alice's subsystem. Even though this shows that quantum steering is sufficient, we also show that quantum correlations are in general not necessary for the conditional generation of Wigner negativity.  

\end{abstract}

\section{Introduction}

Continuous-variable (CV) systems have recently attracted a lot of attention from the quantum computing community, both due to the growing theoretical \cite{Terhal_2020, PhysRevX.8.021054,PRXQuantum.2.020101,PRXQuantum.2.030345, PRXQuantum.2.040353, PRXQuantum.3.010329,PhysRevLett.123.200502} and experimental \cite{ExpErrorCorr1, ExpErrorCorr2, ExpErrorCorr3, ExpErrorCorr4} interest for bosonic codes in quantum error correction, and because of protocols such as Gaussian boson sampling \cite{PhysRevLett.119.170501,PhysRevA.98.062322,PRXQuantum.3.010306,doi:10.1126/sciadv.abl9236,doi:10.1126/science.abe8770,PhysRevLett.127.180502,Borealis}. Such systems are described using observables with a continuous spectrum of possible measurement outcomes, such as position and momentum in mechanical systems, or quadratures of bosonic fields. Even though the physical realisations of such system can vary widely, they are all described within a theoretical framework based on wave functions or phase space representations that distinguishes itself from the discrete variable approach based on qubits or qudits \cite{RevModPhys.84.621,RevModPhys.92.035005,PRXQuantum.2.030204}.

In the CV setting, Gaussian states have been studied in great detail \cite{RevModPhys.84.621}. They can be described using only the state's mean field and covariance matrix, leading to a theoretical description based on symplectic matrix analysis. In experiments Gaussian states are useful because they correspond to the class of states that appear naturally in linear bosonic systems. For example, in quantum optics, Gaussian states coincide nicely to the type of states that can be generated on demand \cite{Su:12,cai-2017,Asavanant:2019aa,Larsen:2019aa}. Yet, in the context of quantum computing, one needs to be able to leave the set of Gaussian states and explore the full state space. In particular, it has been argued that negativity of the Wigner function (one of the typical representations of quantum states on phase space) key for reaching a quantum computation advantage in sampling problems \cite{PhysRevLett.109.230503,Veitch_2012}. Furthermore, it was recently demonstrated that on top some notion of non-Gaussian entanglement, present in any mode basis, is essential for making sampling problems hard to simulate efficiently with classical means \cite{CW2022}.

Wigner negativity and entanglement have thus become relevant resources to study from the point of view of resource theories and quantum state engineering. In previous work, we explored the required resources to remotely generate Wigner negativity in a bipartite system (we refer to the parties as Alice and Bob). If Alice and Bob initially share a Gaussian state, it was shown that a measurement on Bob's subsystem can herald Wigner negativity in Alice's subsystem if and only if Alice can perform Gaussian steering on Bob's subsystem \cite{PhysRevLett.124.150501,PRXQuantum.1.020305,Xiang2022}. These findings have also been demonstrated in photon subtraction experiments \cite{Ra2020,PhysRevLett.128.200401}. These results seemingly interconnects different non-classical resources in a fundamental way. However, it strongly relies on the fact that Alice and Bob initially share a Gaussian state. It remains an open question whether such a connection between the Wigner negativity and quantum steering can be established also for non-Gaussian initial states. In this work, we answer this question.

First of all, we must delineate the exact problem. Since we are interested in a context of remote creation of Wigner negativity in Alice's subsystem, we assume that initially no Wigner negativity is present in Alice's reduced Wigner function. Beyond that, we make no additional assumptions on the global state, nor on the reduced state for Bob's subsystem. Bob will subsequently perform a measurement on his subsystem and Alice will condition on one of Bob's measurement outcomes. In this setting, our most general result of this paper is that quantum steering from Alice to Bob with Wigner positive measurements is a sufficient condition for the remote creation of Wigner negativity, but not a necessary one. 

We start out the article by setting the state in Section \ref{sec:stage}, where we provide a quick introduction to CV systems, the framework of quantum measurements and Wigner functions, and finally quantum steering and associated local hidden state models. In Section \ref{sec:ConditionalSection} we introduce the conditional Wigner function as a useful tool for studying both generation of Wigner negativity and quantum correlations. We also devote some attention to reviewing how this function impact Gaussian steering. In Section \ref{sec:SteeringandCond}, we explore how quantum steering imposes unphysical properties on this conditional Wigner function, but we also show that the presence of these unphysical properties does not necessarily mean there is quantum steering. In Section \ref{sec:negativity}, these unphysical properties of the conditional Wigner function are shown to be responsible for remote generation of Wigner negativity. We thus use the conditional Wigner function as a tool to show that quantum steering is a sufficient but not a necessary resource for the remote generation of Wigner negativity. Finally, we conclude the Article in Section \ref{sec:Conclusions} with a reflection on the relation between Wigner negativity and quantum steering, and we contemplate about the need for a notion of non-local Wigner negativity.

\section{Setting the stage}\label{sec:stage}

\subsection{Continuous-variable systems and the Wigner function}

In this Article, we focus in CV systems \cite{RevModPhys.84.621,RevModPhys.92.035005,PRXQuantum.2.030204}. From a theory point of view, such systems can be understood as ensembles of quantum harmonic oscillators. Their physical realisations come in many forms, ranging from actual mechanical oscillators, over trapped ions, to optics. All of these systems come equipped with a set of observables $\{\hat q_1, \hat p_1, \dots, \hat q_m, \hat p_m\}$, assuming there are $m$ oscillators in the system. These observables behave as the position and momentum operators of the oscillators, and they satisfy the canonical commutation relation $[\hat q_j, \hat p_k] = 2i \delta_{j,k}.$ This notably leads to the Heisenberg inequality ${\rm Var}[\hat q_j]{\rm Var}[\hat p_k] \geq \delta_{jk}$. In quantum optics, the quantum harmonic oscillators describe optical modes, and the ``position'' and ``momentum''  operators describe the {\em quadratures} of the electric field. We will follow this terminology throughout this Article.

It is often convenient to introduce a vector of quandrature operators $\vec{\hat{x}} = (\hat q_1, \hat p_1, \dots, \hat q_m, \hat p_m)^{\top}$ and use it to write the commutation relation as $[\hat x_j, \hat x_k] = 2i \Omega_{jk}$, where we introduce the symplectic structure
\begin{equation}
    \Omega = \bigoplus^m \begin{pmatrix}0 & 1 \\ -1 & 0\end{pmatrix}.
\end{equation}
This matrix also fixes the structure of the phase space that describes the possible outcomes of measurements of quadrature operators. It makes sure that each $\hat q$-operator is paired with a complementary $\hat p$-operator, and any physical operation on the CV system must keep $\Omega$ invariant.

From a mathematical point of view, the quadrature operators are the generators of the algebra of observables. To represent such observables in the phase space of the CV system, we use the Wigner function. The Wigner function \cite{PhysRev.40.749,PhysRev.177.1882,HILLERY1984121} of any observable $\hat A$ can be defined as \cite{PRXQuantum.2.030204}
\begin{equation}\label{eq:Wigner}
    W_{\hat A}(\vec x) = \frac{1}{(2\pi)^m}\int_{\mathbb{R}^{2m}} {\rm tr}[\hat A \, e^{i\vec \alpha^{\top}\vec{\hat{x}}}]e^{-i\vec \alpha^{\top}\vec x}d\vec \alpha.
\end{equation}
In this work, we will consider bipartite systems, where one set of modes is referred to as Alice's system and the other set is known as Bob's system. This structure can be reflected by a direct sum structure on the phase space, such that $\vec x = \vec x_A \oplus \vec x_B$.

The Wigner function of a quantum state $\hat \rho$ is the closest that we can get to a probability distribution on phase space for a quantum state, in the sense that it correctly reproduces all the measurable quadrature statistics as its marginals. Generally speaking, the Wigner function is not always a probability distribution because it can take negative values. However, here we will put our emphasis in the first place on positive Wigner functions, which have all the mathematical properties of a probability distribution on phase space.

\subsection{Quantum measurements}

Wigner functions describing quantum measurements will be a crucial complement to those that describe quantum states. In a general quantum physics context, we describe quantum measurements in terms of positive operator-valued measures $\{\hat \Pi_a \}$ such that ${\rm tr}[\hat \rho \hat \Pi_a]$ quantifies the probability of obtaining a measurement outcome $a$, given that the system was prepared in a state $\hat \rho$. Because the probabilities for all possible measurement outcomes must sum up to one, we must impose the completeness relation \cite{holevo_statistical_2001}
\begin{equation}\label{eq:resId}
    \int_{\cal A} \hat \Pi_a \,da = \mathds{1},
\end{equation}
where ${\cal A}$ is the (possibly continuous) set of measurement outcomes. The $\{\hat \Pi_a \}$ are now said to be a resolution of the identity.

With \eqref{eq:Wigner} we can calculate Wigner functions for these $\hat \Pi_a$, which we will refer to with the shorthand $W_a(\vec x)$. In our work, we typically consider measurements that act on a subsystem. Therefore, when $\hat \Pi_a$ is a measurement of Alice's subsystem, we write its Wigner function as $W_a(\vec x_A)$. The completeness relation \eqref{eq:resId} now translates to an identity for these Wigner functions (see \cite{PRXQuantum.2.030204} for a more complete introduction)
\begin{equation}\label{eq:Completeness}
    (4 \pi)^m\int_{\cal A} W_a(\vec x) \,da = 1 \quad \text{for all } \vec x \in \mathbb{R}^{2m}.
\end{equation}
Where we use that the Wigner function of $\mathds{1}$ is given by $1/(4\pi)^m$. This completeness relation will play an important role in what is to come.

\subsection{Quantum steering}

The framework of quantum steering was developed to describe a type of quantum correlation in which measurements of one subsystem allow to predict measurement results in a correlated subsystem better than classically possible \cite{Schrod1935,Schrod1936,Reid2009,Reid2009Colloquium,Wiseman2007}. This colloquial statement is formalised through a local hidden state (LHS) model that is reminiscent of the local hidden variable model that is used to describe Bell non-locality \cite{Cavalcanti2016,Uola2020}. However, where in the case of Bell inequalities we do not make any assumptions on the local statistics in the model, we do impose some constraints in the case of quantum steering. When we again refer to the two subsystems as Alice and Bob, and we moreover assume that Alice is steering Bob, we will assume that Bob's local statistics is governed by quantum mechanics. No such assumptions are made on Alice's side.

More specifically, Alice will make measurements of some generalised observable $\hat A$ and communicate the measurement outcomes $a$ to Bob. Bob, in turn, will condition his state on these measurement outcomes. If we describe this procedure in the framework of quantum mechanics, we would find that Alice and Bob share an initial state $\hat \rho$ and Alice's measurement means that we apply a POVM element $\hat \Pi_a \otimes \mathds{1}.$ As such, we find that Bob's conditional state is given by
\begin{equation}
    \hat \rho^B_{a \lvert \hat A} = \frac{{\rm tr}_A[(\hat \Pi_a \otimes \mathds{1}) \hat \rho]}{{\rm tr}[(\hat \Pi_a \otimes \mathds{1} )\hat \rho]}, 
\end{equation}
where ${\rm tr}_A[.]$ denoted the partial trace over Alice's subsystem. 

In the context of quantum steering, however, we do not make any assumptions on Alice's subsystem. For all we know, Alice could use some involved computer simulation to generate results that she communicated to Bob. Bob's information is limited to what observable Alice claims to have measured, and which measurement statistics she acquired. We often use this statistics to reformulate the problem in terms of an assemblage, given by
\begin{equation}\label{eq:Assemblage}
    {\cal B}(a, \hat A) = P(a\lvert\hat A)\hat \rho^B_{a \lvert \hat A},
\end{equation}
where we re-scale Bob's state with the probability $P(a\lvert\hat A)$ that Alice finds measurement outcome $a$ upon measuring observable $\hat A$. Bob then wants to test whether this assemblage can be described by the LHS model
\begin{equation}\label{eq:LHS}
   {\cal B}(a, \hat A)  = \int_{\Lambda}P(\lambda) P(a\lvert\hat A,\lambda) \hat \sigma_{\lambda} d \lambda.
\end{equation}
Here we introduce the local hidden variable $\lambda$ that labels a set of local states. The probabilities $P(a\lvert\hat A,\lambda)$ given the chance of obtaining a measurement outcomes $a$ for $\hat A$, given the value of the hidden variable is $\lambda$. These probabilities satisfy $\int_{\Lambda} P(\lambda)P(a\lvert\hat A, \lambda) d\lambda = P(a\lvert\hat A)$. 

Alice is said to be able to steer Bob's measurements if for some observable $\hat A$ the assemblage cannot be written according to \eqref{eq:LHS}. A series of tests, ranging from inference inequalities \cite{Reid2009,PhysRevA.87.062103} and metrological witnesses \cite{YFG21,PRXQuantum.3.030347} to semi-definite programs \cite{Cavalcanti2016,Laurat2018}, have been developed to test whether quantum states satisfy \eqref{eq:LHS} or not. However, in this work, we explore more fundamental connections between the phase space formulation of bosonic quantum systems using the Wigner function and the framework of quantum steering. It is therefore useful to construct a Wigner representation of the assemblage \eqref{eq:Assemblage}:
\begin{equation}\begin{split}\label{eq:AssamblageWigner}
    &W_{\cal B}(\vec x_B) = (4\pi)^l\int_{\mathbb{R}^{2l}} W_a(\vec x_A) W_{\hat \rho}(\vec x_A \oplus \vec x_B)d\vec x_A,
    \end{split}
\end{equation} 
where we assume that Alice's subsystem consists of $l$ modes. 

In the next section, we introduce the conditional Wigner function, the central mathematical object that will forge the connection between Wigner negativity and quantum steering.

\section{Conditional Wigner function}\label{sec:ConditionalSection}

\subsection{General concept}
The definition of the conditional Wigner function is inspired by conditional probability distributions. This object can only be well defined when the Wigner function of at least one of the subsystems is positive (here we assume it to be Alice's), i.e.~$W_{\hat \rho_A}(\vec x_A) > 0$ for all $\vec x_A \in \mathbb{R}^{2l}$. For this specific case we can define the conditional Wigner function \cite{PRXQuantum.1.020305}
\begin{equation}\label{eq:CondWigDef}
    W_{\hat \rho}(\vec x_B \lvert \vec x_A) = \frac{W_{\hat \rho}(\vec x_A \oplus \vec x_B)}{W_{\hat \rho_A}(\vec x_A)}
\end{equation}
Generally speaking, this conditional Wigner function has no specific physical meaning. In classical probability theory, the conditional probability represents the probability for Bob to obtain a certain point $\vec x_B$ in phase space, given that Alice sampled a value $\vec x_A$. However, it is physically impossible to perform any measurement that projects on a specific point in phase space. A priori, this makes the conditional Wigner function a purely mathematical construct. In particular, the conditional Wigner function does not necessarily have all the properties of a well-defined Wigner function of a quantum state for all choices of $\vec x_A$.

To make this statement more formal, we consider a function to be a well-defined Wigner function when it leads to positive expectation values for all positive semi-definite operators $\hat P = \hat X^{\dag} \hat X$ on the Hilbert space of Bob's subsystem. We can use \eqref{eq:Wigner} to construct the Wigner function $W_{\hat P}(\vec x_B)$ of any such positive semi-definite operator and we shall refer to the conditional Wigner function as {\em physical} if and only if
\begin{equation}\label{eq:Physical}
    \int_{\mathbb{R}^{2l'}} W_{\hat P}(\vec x_B)W_{\hat \rho}(\vec x_B \lvert \vec x_A) d \vec x_B \geqslant 0 \text{ for all $\hat P$}.
\end{equation}
As we will see, not all conditional Wigner functions are physical, and thus there are $W_{\hat \rho}(\vec x_B \lvert \vec x_A)$ for which \eqref{eq:Physical} is violated for some choice of $\hat P$.

\subsection{Gaussian states}\label{sec:Gaussian}
A notable example is found for the case of Gaussian states. Any Gaussian state can be represented by a Wigner function of the following form \cite{RevModPhys.84.621}
\begin{equation}\label{eq:Gaussian}
    W_{\hat \rho}(\vec x) = \frac{\exp\left[- \frac{1}{2}(\vec x - \vec \xi)^{\top}V^{-1}(\vec x - \vec \xi)\right]}{(2\pi)^m \sqrt{\det V}},
\end{equation}
where $\vec \xi$ denotes and $m$-dimensional vector that describes the mean field and $V$ is the $m \times m$ covariance matrix. In our bipartite system composed out of Alice and Bob, we structure these objects as $m = l +l'$, $\vec \xi = \vec \xi_A \oplus \vec \xi_B$ and 
\begin{equation}
    V = \begin{pmatrix} V_A & V_{AB} \\ V_{BA} & V_B
    \end{pmatrix}.
\end{equation}
Where $V_A$ ($V_B$) contains the covariances within Alice's (Bob's) subsystem, and $V_{AB} = V_{BA}^{\top}$ describes the correlations between Alice and Bob. The Wigner function \eqref{eq:Gaussian} has all the properties of a Gaussian probability distribution and therefore we know from classical statistics that the conditional Wigner function is given by \cite{PRXQuantum.1.020305}
\begin{equation}\label{eq:ConditionalGaussian}\begin{split}
    &W_{\hat \rho}(\vec x_B \lvert \vec x_A) = \frac{\exp\left[- \frac{1}{2}(\vec x_B - \vec \xi_{B\lvert A})^{\top}V_{B\lvert A}^{-1}(\vec x_B - \vec \xi_{B\lvert A})\right]}{(2\pi)^{l'} \sqrt{\det V_{B\lvert A}}}
    \end{split}
\end{equation}
where the mean field vector is given by 
\begin{equation}
\vec \xi_{B\lvert A} = \vec \xi_B + V_{BA}V_A^{-1}(\vec x_A - \vec \xi_A),    
\end{equation}
and the covariance matrix is given by
\begin{equation}
    V_{B\lvert A} = V_B - V_{BA}V^{-1}_AV_{AB}.
\end{equation}
The latter is known as the Schur complement of $V$ in mathematics, and is guaranteed to be positive whenever the matrix $V$ is a positive matrix. However, the fact that $V$ satisfies the Heisenberg relation is not a guarantee that $V_{B\lvert A}$ will also satisfy this relation. 

In the context of \eqref{eq:Physical}, one can consider the number operator in one of Bob's modes associated with the axis $\vec f \in \mathbb{R}^{2l}$ of Bob's phase space 
\begin{equation}
    \hat n(\vec f) = \frac{1}{4} (\vec f^{\top} \vec{ \hat  x}_B^2 + \vec f^{\top}\Omega \vec{ \hat  x}_B^2 - 2).
\end{equation}
This number operator $\hat n(\vec f)$ is a positive semi-definite operator with Wigner function
\begin{equation}\label{eq:WigNumber}
    W_{\hat n(\vec f)}(\vec x_B) = \frac{1}{16 \pi} (\vec f^{\top} \vec{x}_B^2 + \vec f^{\top}\Omega \vec{x}_B^2 - 2).
\end{equation}
We can now define a general class of positive operators, given by displaced number operators $D(\vec \xi_{B\lvert A} )\hat n(\vec f)D^{\dag}(\vec \xi_{B\lvert A} )$ with Wigner functions
\begin{equation}\label{eq:WigNumberDisp}
    W_{D(\vec \xi_{B\lvert A} )\hat n(\vec f)D^{\dag}(\vec \xi_{B\lvert A} )} (\vec x_B)=  W_{\hat n(\vec f)}(\vec x_B-\vec \xi_{B\lvert A})
\end{equation}
When we insert \eqref{eq:ConditionalGaussian} and \eqref{eq:WigNumberDisp} into \eqref{eq:Physical}, and subsequently work out the Gaussian integrals, we find that for every point $\vec x_A$ in Alice's phase space
    \begin{equation}\begin{split}
    &(4\pi)^{l'}\int_{\mathbb{R}^{2l'}} W_{D(\vec \xi_{B\lvert A} )\hat n(\vec f)D^{\dag}(\vec \xi_{B\lvert A} )}(\vec x_B)W_{\hat \rho}(\vec x_B \lvert \vec x_A) d \vec x_B = \frac{1}{4}( \vec f^{\top} [V_{B\lvert A}+ \Omega^{\top} V_{B\lvert A}\Omega]\vec f- 2)
    \end{split}
\end{equation}
Thus, $W_{\hat \rho}(\vec x_B \lvert \vec x_A)$ is unphysical whenever there is a phase space axis $\vec f$ in Bob's phase space for which $\vec f^{\top} [V_{B\lvert A} + \Omega^{\top} V_{B\lvert A}\Omega]\vec f < 2$. The latter is equivalent to stating that $V_{B\lvert A}$ does not satisfy the uncertainty relation.


For Gaussian states, the physicality of the conditional Wigner function thus depends entirely on the Schur complement $V_{B\lvert A}$. This establishes an intimate relationship with the quantum correlations in the Gaussian state, which can also be studied using $V_{B\lvert A}$ \cite{PhysRevLett.117.220502,Lami_2018}. Most intriguingly, it has been established that Alice can steer Bob with Gaussian measurements if $V_{B\lvert A}$ does not satisfy the uncertainty relation \cite{Wiseman2007,PhysRevLett.114.060403,PhysRevLett.115.210401}. As such, we find that there is a direct link between Gaussian steering and the physicality of the conditional Wigner function. In turn, this has also helped establish a relation between Gaussian steering and the capability of remotely generating Wigner negativity in Alice's subsystem by conditioning on measurement outcomes that occur on Bob's side \cite{PRXQuantum.1.020305}. 

When we leave the realm of Gaussian states and generalise these notions to more general non-Gaussian states, it is unclear to what extend the connections between these different concepts still apply. In what follows, we establish a connection between the physicality of the conditional Wigner function and certain types of quantum steering. However, contrary to the case of Gaussian steering, we find that there is no strict equivalence between steering and physicality of conditional Wigner function.

\section{Quantum steering and the conditional Wigner functions}\label{sec:SteeringandCond}

\subsection{Einstein-Podolsky-Rosen paradox}

First, we study the specific case of steering through homodyne measurements, which literature often refers to as the EPR paradox \cite{EPR1935,PhysRevA.40.913,Reid2009,Reid2009Colloquium}. This paradox is typically expressed in terms of inequalities for conditional variances of homodyne measurements. 

To formalise this, we can select the quadratures that Alice and Bob are going to measure by picking axes $\vec f \in \mathbb{R}^{2l}$ in Bob's phase space and an axis $\vec g \in \mathbb{R}^{2l'}$ in Alice's phase space. We then define operators
\begin{align}
    \hat q_B = \vec f^{\top} \vec{ \hat  x}_B,\\
    \hat p_B = \vec f^{\top} \Omega \vec{ \hat  x}_B,\\
    \hat q_A = \vec g^{\top} \vec{ \hat  x}_A,\\
    \hat p_A = \vec g^{\top} \Omega \vec{ \hat  x}_A.
\end{align}
First of all, we defined the conditional probability 
\begin{equation}\begin{split}
    &P(q_B \lvert q_A)= \frac{\int_{\mathbb{R}^{2m}} \delta(\vec f^{\top} \vec{x}_B - q_B)\delta(\vec g^{\top} \vec{  x}_A - q_A) W_{\hat \rho}(\vec x_A \oplus \vec x_B) d\vec x_A d\vec x_B}{\int_{\mathbb{R}^{2m}}\delta(\vec g^{\top} \vec{x}_A - q_A) W_{\hat \rho}(\vec x_A \oplus \vec x_B) d\vec x_A d\vec x_B }.
    \end{split}
\end{equation}
The conditional probability $P(p_B \lvert p_A)$ for the complementary quadratures is defined analogously. We can now use the conditional probability to defined its variance
\begin{equation}
{\rm Var}[\hat q_B | \hat q_A = q_A] = \int_{\mathbb{R}} q_B^2 P(q_B \lvert q_A)dq_B - \left(\int_{\mathbb{R}} q_B P(q_B \lvert q_A)dq_B\right)^2. 
\end{equation}
Again the variance ${\rm Var}[\hat p_B \mid \hat p_A = p_A]$ of $P(p_B \lvert p_A)$ is defined analogously. These variances can depend strongly on the choice of measurement outcome $q_A$ and $p_A$, it therefore more useful to define the conditional variance 
\begin{equation}
    {\rm Var}[\hat q_B | \hat q_A] = \int_{\mathbb{R}} P(q_A){\rm Var}[\hat q_B |\hat q_A = q_A],
\end{equation}
where we average the variance of the conditional probability distribution of the probability to obtain the outcome that is conditioned upon.

In literature these conditional variances are used as steering witnesses in what is typically referred to as Reid's criterion \cite{PhysRevA.40.913,Reid2009}:
\begin{equation}
    {\rm Var}[\hat q_B \lvert \hat q_A]{\rm Var}[\hat p_B \lvert \hat p_A] < 1 \implies \text{quantum steering}.
\end{equation}
For Gaussian states, these conditional variances take a specifically simple form \cite{PhysRevA.40.913} and, on top, they are upper bounds for the components of the Schur complement $V_{B\lvert A}$ \cite{PRXQuantum.2.030204}. For non-Gaussian states, the Schur complement of the covariance matrix can still be used to detect quantum steering \cite{PhysRevLett.115.210401}, but it is no longer directly related to the conditional variances of homodyne detection. Yet, we will now show that ${\rm Var}[\hat q_B \lvert \hat q_A]$ is still an upper-bound for the variance of the conditional Wigner function $W_{\hat \rho}(\vec x_B \lvert \vec x_A)$.
 
To show this, we first define the probability to obtain a measurement outcome $q_A$ as
\begin{equation}
    P(q_A) = \int_{\mathbb{R}^{2l'}}\delta(\vec g^{\top} \vec{x}_A - q_A) W_{\hat \rho}(\vec x_A \oplus \vec x_B)d\vec x_A d\vec x_B.
\end{equation}
We can now write that
\begin{align}
    W_{\hat \rho}(\vec x_B \lvert q_A) &= \frac{\int_{\mathbb{R}^{2l'}}\delta(\vec g^{\top} \vec{ x}_A - q_A) W_{\hat \rho}(\vec x_A \oplus \vec x_B)d\vec x_A}{P(q_A)}\\
    &= \int_{\mathbb{R}^{2l'}}\delta(\vec g^{\top} \vec{ x}_A - q_A)\frac{ W_{\hat \rho}(\vec x_A \oplus \vec x_B)}{W_{\hat \rho}(\vec x_A)}\frac{W_{\hat \rho}(\vec x_A)}{P(q_A)}d\vec x_A \label{eq:ConditionalIdentity1}
\end{align}
To make notation more explicit, it is now useful go to a specific basis of Alice's phase space, such that $\vec x_A = (q_A,p_A, q^{(2)}_A, p^{(2)}_A, \dots, q^{(l')}_A, p^{(l')}_A)^{\top}$. Then we can rewrite \eqref{eq:ConditionalIdentity1} to find
\begin{align}\label{eq:conditionalIdentity2}
    W_{\hat \rho}(\vec x_B \lvert q_A)
    = \int_{\mathbb{R}^{2l'-1}}& W_{\hat \rho}(\vec x_B | q_A,p_A, q^{(2)}_A, p^{(2)}_A, \dots, q^{(l')}_A, p^{(l')}_A) \\
    &\nonumber \times W_{\hat \rho}(p_A, q^{(2)}_A, p^{(2)}_A, \dots, q^{(l')}_A, p^{(l')}_A| q_A )\,dp_A q^{(2)}_A d p^{(2)}_A \dots d q^{(l')}_A d p^{(l')}_A,
\end{align}
 where the first factor in the integral is simply the conditional Wigner function $W_{\hat \rho}(\vec x_B \lvert \vec x_A)$ writing in the specific basis of Alice's phase space.

As a next step, let us define the averaged variance of the conditional Wigner function as
\begin{equation}\label{eq:averageVarianceCondWig}
    {\rm Var}_c[q_B] = \int_{\mathbb{R}^{2l'}}W_{\hat \rho}(\vec x_A) {\rm Var}[q_B | \vec x_A],
\end{equation}
with the variance of the conditional Wigner function given by
\begin{equation}
    {\rm Var}[q_B | \vec x_A] = \int_{\mathbb{R}^{2l}} (\vec f^{\top} \vec{ x}_B)^2 W(\vec x_B|\vec x_A)d\vec x_B - \left(\int_{\mathbb{R}^{2l}} \vec f^{\top} \vec{  x}_B W(\vec x_B|\vec x_A)d\vec x_B\right)^2.
\end{equation}
Notice that  \eqref{eq:averageVarianceCondWig} makes sense as an averaged variance only because we assume that Alice's Wigner function $W_{\hat \rho}(\vec x_A)$ is positive. Of course, ${\rm Var}_c[p_B]$ and $ {\rm Var}[p_B | \vec x_A]$ can be defined accordingly.

In what follows, we show that ${\rm Var}[\hat q_B \lvert \hat q_A] \geq {\rm Var}_c[q_B]$ and analogously ${\rm Var}[\hat p_B \lvert \hat p_A] \geq {\rm Var}_c[p_B]$. To do so, we can start by the first term in ${\rm Var}[\hat q_B \lvert \hat q_A]$, given by
\begin{equation}
    \int_{\mathbb{R}} P(q_A) \int_{\mathbb{R}} q_B^2 P(q_B \lvert q_A)dq_B dq_A = \int_{\mathbb{R}^{2m}} (\vec f^{\top} \vec{ \hat  x}_B)^2 W_{\hat \rho}(\vec x_A \oplus \vec x_B) d\vec x_A d\vec x_B.
\end{equation}
The first term in ${\rm Var}_c[q_B]$ yields the same expression
\begin{equation}\begin{split}
    \int_{\mathbb{R}^{2l'}}W(\vec x_A) \int_{\mathbb{R}^{2l}}(\vec f^{\top} \vec{ x}_B)^2 W(\vec x_B|\vec x_A) d\vec x_A d\vec x_B &= \int_{\mathbb{R}^{2m}} (\vec f^{\top} \vec{  x}_B)^2 W_{\hat \rho}(\vec x_A \oplus \vec x_B) d\vec x_A d\vec x_B
    \end{split}
\end{equation}
This shows that the only difference between ${\rm Var}[\hat q_B \lvert \hat q_A]$ and ${\rm Var}_c[q_B]$ terms comes from the second terms. To see this difference, we can write that
\begin{equation}\begin{split}
   \int_{\mathbb{R}} &P(q_A)\left(\int_{\mathbb{R}} q_B P(q_B \lvert q_A)dq_B\right)^2dq_A \\
   &=  \int_{\mathbb{R}} P(q_A)\left(\int_{\mathbb{R}^ {2l}} \vec f^{\top} \vec{ x}_B  W(\vec x_B \lvert q_A)d\vec x_B\right)^2dq_A \\
   &\stackrel{\eqref{eq:conditionalIdentity2}}{=}  \int_{\mathbb{R} }P(q_A)\Bigg(\int_{\mathbb{R}^ {2l}} \vec f^{\top} \vec{ x}_B  \int_{\mathbb{R}^{2l'-1}} W_{\hat \rho}(\vec x_B | q_A,p_A, q^{(2)}_A, p^{(2)}_A, \dots, q^{(l')}_A, p^{(l')}_A) \\
    &\qquad\qquad \times W_{\hat \rho}(p_A, q^{(2)}_A, p^{(2)}_A, \dots, q^{(l')}_A, p^{(l')}_A| q_A )\,dp_A q^{(2)}_A d p^{(2)}_A \dots d q^{(l')}_A d p^{(l')}_Ad\vec x_B\Bigg)^2dq_A \\
    &\leq \int_{\mathbb{R}^{2l} }W_{\hat \rho_A}(\vec x_A)\Bigg(\int_{\mathbb{R}^ {2l}} \vec f^{\top} \vec{x}_B  \int_{\mathbb{R}^{2l'-1}} W_{\hat \rho}(\vec x_B | \vec x_A)d\vec x_B\Bigg)^2d\vec x_A,
   \end{split}
\end{equation}
where in the final step we used Jensen's inequality. The above inequality has as a direct consequence that ${\rm Var}[\hat q_B \lvert \hat q_A] \geq{\rm Var}_c[q_B]$. Because we can repeat all the same steps for the $p$-quadratures, we find that
\begin{equation}
    {\rm Var}[\hat q_B \lvert \hat q_A]{\rm Var}[\hat p_B \lvert \hat p_A] \geq {\rm Var}_c[q_B]{\rm Var}_c[p_B].
\end{equation}
As a consequence, when there is quantum steering with homodyne measurements, and thus ${\rm Var}[\hat q_B \lvert \hat q_A]{\rm Var}[\hat p_B \lvert \hat p_A] < 1$, we find that also ${\rm Var}_c[q_B]{\rm Var}_c[p_B] < 1$. 

Finally, we can apply the Cauchy-Schwarz inequality to show that when ${\rm Var}_c[q_B]{\rm Var}_c[p_B] < 1$, there exists at least one $\vec x_A$ for which $ {\rm Var}[q_B | \vec x_A] {\rm Var}[p_B | \vec x_A] < 1$. This concludes that 
\begin{equation}
    {\rm Var}[\hat q_B \lvert \hat q_A]{\rm Var}[\hat p_B \lvert \hat p_A] < 1 \implies {\rm Var}[q_B | \vec x_A] {\rm Var}[p_B | \vec x_A] < 1 \text{ for some $\vec x_A \in \mathbb{R}^{2l'}$},
\end{equation}
and thus steering with homodyne measurements implies that $W(\vec x_B | \vec x_A)$ is unphysical. In this sense, this is a direct generalisation of the Gaussian case. However, note that the inverse implication does not hold. In other words, an unphysical conditional Wigner function does not automatically cause an EPR paradox.

A natural question is whether these finding generalise to steering via other kinds of measurements, possibly beyond the Gaussian realm. In the next section, we will consider the case of steering by Wigner positive measurements and provide an affirmative answer to this question.

\subsection{Quantum steering with Wigner-positive measurements}\label{sec:sufficient}

Going back to \eqref{eq:Completeness}, we can define Wigner functions associated with the POVM elements that describe quantum measurements. In the previous section, we focused on the case where steering manifests by homodyne measurements on Alice's side and can be witnessed through the conditional variance of Bob's homodyne measurements. In terms of POVM elements, this means that the steering is done with Wigner functions of the form $W_q(\vec x_A) \propto \delta (\vec g^{\top} \vec{ \hat  x}_A - q_A)$ and $W_p(\vec x_A) \propto \delta (\vec g^{\top}\Omega \vec{ \hat  x}_A - p_A)$. These are of course highly specific measurements, which we will now generalise to the case where the Wigner functions that describe Alice's POVM elements are just assumed to be positive. In other words, Alice can perform any measurement that satisfies \eqref{eq:Completeness}, provided that $W_a(\vec x_A)$ is strictly positive for all possible measurement outcomes $a$.

We start the technical analysis of this scenario by recasting the Wigner function of the assemblage \eqref{eq:AssamblageWigner} in the following form:
\begin{equation}\begin{split}\label{eq:AssamblageWigner2}
    &W_{\cal B}(\vec x_B) = (4\pi)^l\int_{\mathbb{R}^{2l}} W_a(\vec x_A) W_{\hat \rho_A}(\vec x_A) W_{\hat \rho}(\vec x_B \lvert \vec x_A)  d\vec x_A.
    \end{split}
\end{equation}
This suggests a connection to the LHS model \eqref{eq:LHS} with $\vec x_A$ as hidden variable. To make this connection explicit, we recast \eqref{eq:LHS} in a phase space form
\begin{equation}\label{eq:LHSWigner}
    W_{\cal B}(\vec x_B) = \int_{\Lambda}P(\lambda) P(a\lvert\hat A,\lambda) W_{\lambda}(\vec x_B) d \lambda,
\end{equation}
where we introduce $W_{\lambda}(\vec x_B)$ as a shorthand for the Wigner function of $\hat \sigma_{\lambda}$. Recall that, throughout the whole Article, we assume explicitly that $\hat \rho_A$ has a positive Wigner function, and thus we can identify $W_{\hat \rho_A}(\vec x_A)$ with a probability distribution. Thus, we can recast $W_{\hat \rho_A}(\vec x_A) \mapsto P(\vec x_A)$, to indicated the probability density of obtaining a phase space coordinate $\vec x_A$.

We then need to use the completeness relation \eqref{eq:Completeness} for the measurements of Alice's observable $\hat A$. In general terms, the Wigner functions $W_a(\vec x_A)$ could have negative regions, which makes the relation \eqref{eq:Completeness} not very useful. However, for the particular case of measurements with positive Wigner functions, we find that, regardless of $\vec x_A$, the values $(4\pi)^lW_a(\vec x_A)$ are positive for all $a$, and they sum  up to one. This means that we can identify $(4\pi)^lW_a(\vec x_A) \mapsto P(a\lvert \hat A, \vec x_A)$, the probability of finding an outcome $a$, given measurement setting $\hat A$ and phase space coordinate $\vec x_A$. In this regard, we find that \eqref{eq:AssamblageWigner2} becomes
\begin{equation}\begin{split}\label{eq:ThisWigner}
      &W_{\cal B}(\vec x_B) =\int_{\mathbb{R}^{2l}} P(\vec x_A) P(a \lvert \hat A, \vec x_A) W_{\hat \rho}(\vec x_B \lvert \vec x_A)  d\vec x_A.
    \end{split}
\end{equation}
We now see that the steerability depends on the properties of $W_{\hat \rho}(\vec x_B \lvert \vec x_A)$. More specifically, we find that if $W_{\hat \rho}(\vec x_B \lvert \vec x_A)$ is actually a well-defined Wigner function of a quantum state for every $\vec x_A$, the LHS model \eqref{eq:LHSWigner} is satisfied with $\vec x_A$ as hidden variable. This leads us to a fundamental conclusion that a physical $W_{\hat \rho}(\vec x_B \lvert \vec x_A)$ for all $\vec x_A$ implies that it is impossible for Alice to steer Bob with Wigner-positive measurements. By contra-position, we find that if Alice can steer Bob's system with Wigner-positive measurements and Alice's local state also has a positive Wigner function, the conditional Wigner function $W_{\hat \rho}(\vec x_B \lvert \vec x_A)$ must be unphysical for some phase space coordinates $\vec x_A$.

\subsection{Unphysical conditional Wigner functions do not imply quantum steering}

Up to here, we have explored the consequences of quantum steering with Wigner-positive measurements on the conditional Wigner function $W_{\hat \rho}(\vec x_B \lvert \vec x_A)$. However, it is natural to wonder whether some type of reverse statement is possible. In other words, when $W_{\hat \rho}(\vec x_B \lvert \vec x_A)$ is not physical, does it imply steering with Wigner-positive measurements? This statement is equivalent that showing that a LHS model \eqref{eq:LHSWigner} implies that $W_{\hat \rho}(\vec x_B \lvert \vec x_A)$ must be physical when Alice's local state has a positive Wigner function and applies Wigner-positive measurements. There is, however, a clear way to construct counterexamples for this idea.

Let us consider a separable state $\hat \rho$ with a Wigner function given by
\begin{equation}
    W_{\hat \rho}(\vec x_A \oplus \vec x_B) = \sum_{\lambda} p_{\lambda}W_{\hat \rho^{\lambda}_A}(\vec x_A) W_{\hat \rho^{\lambda}_B}(\vec x_B)
\end{equation}
The conditional Wigner function thus becomes
\begin{equation}
    W_{\hat \rho}(\vec x_B \lvert \vec x_A) = \frac{\sum_{\lambda} p_{\lambda}W_{\hat \rho^{\lambda}_A}(\vec x_A) W_{\hat \rho^{\lambda}_B}(\vec x_B)}{\sum_{\lambda} p_{\lambda}W_{\hat \rho^{\lambda}_A}(\vec x_A)},
\end{equation}
and we assume that $\sum_{\lambda} p_{\lambda}W_{\hat \rho^{\lambda}_A}(\vec x_A)$ is a positive function. The condition for physicality now boils down to
\begin{equation}
\begin{split}
    &\int_{\mathbb{R}^{2l'}} W_{\hat P}(\vec x_B)W_{\hat \rho}(\vec x_B \lvert \vec x_A) d \vec x_B   = \frac{\sum_{\lambda} p_{\lambda}W_{\hat \rho^{\lambda}_A}(\vec x_A) \int_{\mathbb{R}^{2l'}} W_{\hat P}(\vec x_B) W_{\hat \rho^{\lambda}_B}(\vec x_B)d\vec x_B}{\sum_{\lambda} p_{\lambda}W_{\hat \rho^{\lambda}_A}(\vec x_A)} < 0,
\end{split}
\end{equation}
for all positive semi-definite operators $\hat P$. Without loss of generality, we can assume that $\hat \rho^{\lambda}_A$ and $\hat \rho^{\lambda}_B$ are pure states for every $\lambda$. In this case, we can select $\hat P = \hat \rho^{\lambda'}_B$ such that
\begin{equation}
    \int_{\mathbb{R}^{2l'}} W_{\hat P}(\vec x_B) W_{\hat \rho^{\lambda}_B}(\vec x_B)d\vec x_B = \delta_{\lambda', \lambda}.
\end{equation}
This implies that the function $W_{\hat \rho}(\vec x_B \lvert \vec x_A)$ is not a physical Wigner function whenever there is a $\lambda'$ such that
\begin{equation}
\frac{ p_{\lambda}W_{\hat \rho^{\lambda'}_A}(\vec x_A)}{\sum_{\lambda} p_{\lambda}W_{\hat \rho^{\lambda}_A}(\vec x_A)} < 0.
\end{equation}
By construction, we assumed that $\sum_{\lambda} p_{\lambda}W_{\hat \rho^{\lambda}_A}(\vec x_A) = W_{\hat \rho_A}(\vec x_A)$ is a positive Wigner function. Thus, the conditional Wigner function is not physical whenever we can find a $\lambda'$ for which
\begin{equation}\label{eq:condPureNonGauss}
   W_{\hat \rho^{\lambda'}_A}(\vec x_A) < 0.
\end{equation}
Because on top we can always find a convex decomposition in which $\hat \rho^{\lambda'}_A$ is a pure state, the condition \eqref{eq:condPureNonGauss} holds for all non-Gaussian states. In other words, any separable states that is not a mixture of Gaussian states will lead to a conditional Wigner function $W_{\hat \rho}(\vec x_B \lvert \vec x_A)$ that is not physical.

This construction provides a general class of counterexamples for which the physicality of $W_{\hat \rho}(\vec x_B \lvert \vec x_A)$ is not equivalent to the possibility to have quantum steering from Alice to Bob. In particular, there are many separable states for which $W_{\hat \rho}(\vec x_B \lvert \vec x_A)$ is unphysical.\\

Going back to the Gaussian case of Section \ref{sec:Gaussian}, one might argue that steerable Gaussian states do not only have a conditional Wigner function that is unphysical for one specific value of $\vec x_A$, but rather one that is unphysical for every possible choice of $\vec x_A$. Yet, even when we take this as a stricter condition, we can construct an explicit counterexample.

We will consider a specific type of separable state for Alice and Bob, given by
\begin{equation}\label{eq:littleCounterExample}
    \hat \rho_{AB} = \sum_{n=0}^{\infty} p_n \ket{n}\bra{n} \otimes \ket{n}\bra{n},
\end{equation}
where $\ket{n}$ denotes a single-mode $n$ photon Fock state. Such a state has a perfect, yet fully classical, correlation between the number of photons in Alice's and Bob's mode. The Wigner function for this separable state is given by
\begin{equation}
   W_{\hat \rho_{AB}}(\vec x_A \oplus \vec x_B) = \sum_{n=0}^{\infty} p_n W_{\ket{n}\bra{n}}(\vec x_A) W_{\ket{n}\bra{n}}(\vec x_B) ,
\end{equation}
and therefore the conditional Wigner function is given by 
\begin{equation}
    W_{\hat \rho_{AB}}(\vec x_B \lvert \vec x_A) = \sum_{n=0}^{\infty} \frac{p_n W_{\ket{n}\bra{n}}(\vec x_A)}{\sum_{n'} p_{n'}  W_{\ket{n'}\bra{n'}}(\vec x_A)} W_{\ket{n}\bra{n}}(\vec x_B).
\end{equation}
For this function to be well-defined, we assume that the probabilities $p_n$ are such that $W_{\hat \rho_A}(\vec x_A) = \sum_{n'} p_{n'}  W_{\ket{n'}\bra{n'}}(\vec x_A) \geq 0$ for all $\vec x_A$. Furthermore, we want the distribution to be such that $p_n > 0$ for all $n$. These requirements can be satisfied by choosing $p_n = t^n/(1+t)^{(n+1)}$ for some parameter $t>0$, such that we obtain the probabilities of a thermal state.

To check the physicality of $W_{\hat \rho_{AB}}(\vec x_B \lvert \vec x_A)$, we will consider a positive operator of the form $\hat P = \ket{m}\bra{m}$. First of all, we use that  
\begin{equation}
    4\pi \int_{\mathbb{R}^{2}} W_{\ket{n}\bra{n}}(\vec x_B) W_{\ket{m}\bra{m}}(\vec x_B) = \delta_{n,m},
\end{equation}
to show that
\begin{equation}
    4\pi \int_{\mathbb{R}^{2}} W_{\hat \rho_{AB}}(\vec x_B \lvert \vec x_A) W_{\ket{m}\bra{m}}(\vec x_B) = \frac{p_m W_{\ket{m}\bra{m}}(\vec x_A)}{\sum_{n'} p_{n'}  W_{\ket{n'}\bra{n'}}(\vec x_A)}.
\end{equation}
Thus, the question is now whether for every point in phase space $\vec x_A$, we can find a number of photons $m$ such that $W_{\ket{m}\bra{m}}(\vec x_A) < 0$. We will prove this possibility by using the properties of Laguerre polynomials. 

The Wigner function of and $m$-photon Fock state is given by 
\begin{equation}
    W_{\ket{m}\bra{m}}(\vec x_A)= (-1)^m L_m(\norm{\vec x_A}^2)\frac{e^{-\frac{\norm{\vec x_A}^2}{2}}}{2\pi},
\end{equation}
where $L_m(\vec x_A)$ is the $m$th Laguerre polynomial. Laguerre polynomials have one particularly useful property that comes in the form of a recurrence relation
\begin{equation}
    (m+1)L_{m+1}(x) = (2m+1 - x) L_m(x) - m L_{m-1}(x).
\end{equation}
This relation can be used to prove the follow identity for the Wigner functions of Fock states
\begin{equation}\label{eq:FockRecurrence}
    (m+1)W_{\ket{m + 1}\bra{m +1 }}(\vec x_A) = (\norm{\vec x_A}^2 - 2m - 1) W_{\ket{m}\bra{m}}(\vec x_A) - m W_{\ket{m-1}\bra{m-1}}(\vec x_A),
\end{equation}
where we used that $(-1)^{m+1} = -(-1)^m = (-1)^{m-1}$. Now let us assume that there is a point $\vec x_A$ in phase space for which $W_{\ket{m}\bra{m}}(\vec x_A) \geq 0$ for all $m$. Under this assumption, the relation \eqref{eq:FockRecurrence} implies that for all $m$
\begin{equation}
    (\norm{\vec x_A}^2 - 2m - 1) W_{\ket{m}\bra{m}}(\vec x_A) \geq m W_{\ket{m-1}\bra{m-1}}(\vec x_A).
\end{equation}
Note that the left-hand side it always positive because of our assumption, which means that $(\norm{\vec x_A}^2 - 2m - 1) W_{\ket{m}\bra{m}}(\vec x_A) \geq 0.$
Because $W_{\ket{m}\bra{m}}(\vec x_A) \geq 0$ for all $m$, the above inequality can only be satisfied when $(\norm{\vec x_A}^2 - 2m - 1) \geq 0$ for all $m$. However, regardless of $\vec x_A$, we can always find some natural number $m$ such that this inequality does not hold. This leads to a contradiction, thus our initial assumption, that a point $\vec x_A$ exists for which $W_{\ket{m}\bra{m}}(\vec x_A) \geq 0$ for all $m$, must be false.

This proves that the existence of a separable state, given by \eqref{eq:littleCounterExample}, with a conditional Wigner function $W_{\hat \rho_{AB}}(\vec x_B \lvert \vec x_A)$ that is unphysical for every point $\vec x_A$ in Alice's phase space. Hence, in general the unphysicality of the conditional Wigner function does not imply quantum steering.

\section{Remote generation of Wigner negativity}\label{sec:negativity}

The conditional Wigner function has been shown to be a crucial element in the study of quantum state preparation. We go to a context where Bob performs measurements on his subsystem and communicates the outcome to Alice. Alice, in turn, will post-select her state on on specific measurement outcome $b$ on Bob's side, corresponding the a POVM element $\hat P_b$. As such, Alice obtains a conditional state given by
\begin{equation}
    \hat \rho_{A\lvert b} = \frac{\tr_B[(\mathds{1}\otimes \hat P_B) \hat \rho_{AB}]}{\tr[(\mathds{1}\otimes \hat P_B) \hat \rho_{AB}]}
\end{equation}
When we translate this expression to phase space, we find \cite{PRXQuantum.1.020305}
\begin{equation}\label{eq:wigCond}
    W_{A\lvert b}(\vec x_A) = \frac{\<\hat P_b\>_{B\lvert \vec x_A}}{\<\hat P_b\>}W_{\hat \rho_A}(\vec x_A),
\end{equation}
which is can be understood as Bayes rule for Wigner functions. We define $\<\hat P_b\>$ simply as 
\begin{equation}
    \<\hat P_b\> = (4\pi)^{l'} \int_{\mathbb{R}^{2l'}}W_{\hat P_b}(\vec x_B)W_{\hat \rho_B}(\vec x_B)d\vec x_B,
\end{equation}
which denotes the probability for Bob to obtain the specific measurement outcome. On the other hand, we also introduce what can be called the conditional quasi-probability, given by
\begin{equation}\label{eq:condQuasiProb}
\<\hat P_b\>_{B\lvert \vec x_A} = (4\pi)^{l'} \int_{\mathbb{R}^{2l'}}W_{\hat P_b}(\vec x_B)W_{\hat \rho_{AB}}(\vec x_B\lvert \vec x_A)d\vec x_B.
\end{equation}
A particular feature of this object, which shares some features with a normal probability, is its capability of attaining negative values for certain $\vec x_A$. This again reflects the fact that it is physically impossible to condition on a well-defined point in phase space. Note that $\<\hat P_b\>_{B\lvert \vec x_A}$ is only well-defined for all $\vec x_A$ when $W_{\hat \rho_A}(\vec x_A)$ is a positive Wigner function. Thus, we can only apply \eqref{eq:wigCond} in cases where Alice's reduced initial state has no Wigner negativity. 

It will be exactly in cases where the conditional quasi-probabilities $\<\hat P_b\>_{B\lvert \vec x_A}$ are negative, that Alice's Wigner function will reach negative values after conditioning on Bob's measurements. In the light of \eqref{eq:Physical} the condition that certain phase space points $\vec x_A$ exist for which $\<\hat P_b\>_{B\lvert \vec x_A} < 0$ is narrowly related to the physicality of $W_{\hat \rho_{AB}}(\vec x_B \lvert \vec x_A)$. If Alice and Bob share a state $\hat \rho_{AB}$ with a conditional Wigner function $W_{\hat \rho_{AB}}(\vec x_B \lvert \vec x_A)$ that is physical, no measurement on Bob's side can create Wigner negativity on Alice's subsystem. However, when we find that $W_{\hat \rho_{AB}}(\vec x_B \lvert \vec x_A)$ is unphysical for some value $\vec x_A$, the combination of \eqref{eq:Physical} and \eqref{eq:condQuasiProb} implies the existence of a positive operator $\hat P_b$ for which $\<\hat P_b\>_{B\lvert \vec x_A} < 0$ for that specific value $\vec x_A$. Therefore, we can create Wigner negativity on Alice's subsystem by letting Bob perform a measurement with POVM element related to this particular positive operator $\hat P_b$.\\

For Gaussian states, we showed in previous work that $W_{\hat \rho_{AB}}(\vec x_B \lvert \vec x_A)$ is physical if and only if Alice cannot steer Bob with Gaussian measurements \cite{PRXQuantum.1.020305}. In Section \ref{sec:Gaussian} we explicitly show that the operator $\hat P$, for which the physicality condition \eqref{eq:Physical} is violated, is the displaced number operator. Thus, we establish a more solid connection to a previous result \cite{PhysRevLett.124.150501} that shows how photon subtraction on Bob's subsystem can create Wigner-negativity on Alice's system is and only if Alice can steer Bob with Gaussian measurements.

Our above analysis shows, however, that quantum steering is no longer a necessary and sufficient condition when we go to non-Gaussian states. The unphysicality of the conditional Wigner function $W_{\hat \rho_{AB}}(\vec x_B \lvert \vec x_A)$ still enables the creation of Wigner-negativity, but an unphysical conditional Wigner function no longer implies quantum steering. Quantum steering from Alice to Bob is thus not necessary for the creation of Wigner negativity in Alice's subsystem. Nevertheless, we did prove that it is sufficient. Indeed, in Section \ref{sec:sufficient} we show that states with a physical conditional Wigner function can never be steered by Wigner-positive measurements and by contra-position, Alice's ability to steer Bob's state with Wigner-positive measurement implies that $W_{\hat \rho_{AB}}(\vec x_B \lvert \vec x_A)$ is unphysical and thus we reach the central conclusion:
\begin{equation}\label{eq:mainResult}
\begin{split}
&\text{Alice can steer Bob with Wigner-positive measurements} \\
&\overset{\centernot{\impliedby}}{\implies} \text{Bob can remotely create Wigner-negativity in Alice's subsystem}.
\end{split}
\end{equation}

\section{Conclusions}\label{sec:Conclusions}

In this article, we started by providing a general definition of the condition Wigner function that can be applied to non-Gaussian and Wigner-negative states. The only constraint on our construction is that Alice must have a reduced state with a positive Wigner function. Generalising a previous result, we can show that Wigner-negativity can be created in Alice's reduced state if and only if the conditional Wigner function is unphysical. We explore whether the physicality of the conditional Wigner function is related to quantum steering, as is known to be the case for Gaussian states. We found that the ability of Alice to steer Bob's measurements with Wigner-positive measurements implies that the conditional Wigner function must me unphysical. However, the inverse statement does not hold, {\it i.e.}, an unphysical conditional Wigner function does not automatically mean there is steering. We illustrate this with counterexamples. Combining these elements leads us to the main result of this Article, given in \eqref{eq:mainResult}: Alice's ability to steer Bob's subsystem with Wigner-positive measurements is a sufficient condition for Bob to be able to remotely create Wigner negativity in Alice's subsystem with some quantum operation. However, it is not a necessary condition.  

The counterexamples presented all impose a clear message: mixtures of states with classically correlated Wigner negativity can be used for the conditional creation of Wigner negativity. Even though the physicality of the conditional Wigner function is a great tool for understanding Wigner negativity, it is less useful for studying quantum correlations. Unphysical conditional Wigner functions can arise from mixtures of states with local Wigner negativity, which becomes evident when we calculate the conditional Wigner functions of separable mixed states with a non-zero stellar rank \cite{PhysRevLett.124.063605,PhysRevLett.106.200401}. Note that there are non-Gaussian mixtures of Gaussian states, which are of stellar rank zero \cite{PRXQuantum.2.030204}. Our counterexamples do not apply to these states, and thus it is still an open question whether the conditional Wigner function can unveil steering properties of such mixtures of Gaussian states. However, because these states are of limited interest for most quantum technologies, we do not treat this problem here.

The findings of this work highlight the need for a genuine notion of nonlocal Wigner negativity in order to understand quantum correlations in non-Gaussian states. It is already established that Wigner negativity in the state is necessary to violate a Bell inequality with Wigner-positive measurements \cite{PRXQuantum.2.030204,Ferrie_2011}. Furthermore, Wigner negativity has been shown to be equivalent to the presence of contextuality in a continuous-variable context \cite{Contextuality1,Contextuality2}. It does seem natural that a notion of nonlocal Wigner negativity, and thus nonlocal contextuality, underlies at least certain types of non-Gaussian quantum correlations. 

\section*{Acknowledgements}
I thank Nicolas Treps and Maria Maffei for fruitful discussions that motivated this work. I also thank Ulysse Chabaud and Federico Centrone for inspiring discussions about the general topic of quantum correlations in CV systems. This work was funded by ANR JCJC project NoRdiC (ANR-21-CE47-0005).

\bibliographystyle{apsrev4-2}
\bibliography{biblio.bib}

\begin{thebibliography}{59}%
\makeatletter
\providecommand \@ifxundefined [1]{%
 \@ifx{#1\undefined}
}%
\providecommand \@ifnum [1]{%
 \ifnum #1\expandafter \@firstoftwo
 \else \expandafter \@secondoftwo
 \fi
}%
\providecommand \@ifx [1]{%
 \ifx #1\expandafter \@firstoftwo
 \else \expandafter \@secondoftwo
 \fi
}%
\providecommand \natexlab [1]{#1}%
\providecommand \enquote  [1]{``#1''}%
\providecommand \bibnamefont  [1]{#1}%
\providecommand \bibfnamefont [1]{#1}%
\providecommand \citenamefont [1]{#1}%
\providecommand \href@noop [0]{\@secondoftwo}%
\providecommand \href [0]{\begingroup \@sanitize@url \@href}%
\providecommand \@href[1]{\@@startlink{#1}\@@href}%
\providecommand \@@href[1]{\endgroup#1\@@endlink}%
\providecommand \@sanitize@url [0]{\catcode `\\12\catcode `\$12\catcode
  `\&12\catcode `\#12\catcode `\^12\catcode `\_12\catcode `\%12\relax}%
\providecommand \@@startlink[1]{}%
\providecommand \@@endlink[0]{}%
\providecommand \url  [0]{\begingroup\@sanitize@url \@url }%
\providecommand \@url [1]{\endgroup\@href {#1}{\urlprefix }}%
\providecommand \urlprefix  [0]{URL }%
\providecommand \Eprint [0]{\href }%
\providecommand \doibase [0]{https://doi.org/}%
\providecommand \selectlanguage [0]{\@gobble}%
\providecommand \bibinfo  [0]{\@secondoftwo}%
\providecommand \bibfield  [0]{\@secondoftwo}%
\providecommand \translation [1]{[#1]}%
\providecommand \BibitemOpen [0]{}%
\providecommand \bibitemStop [0]{}%
\providecommand \bibitemNoStop [0]{.\EOS\space}%
\providecommand \EOS [0]{\spacefactor3000\relax}%
\providecommand \BibitemShut  [1]{\csname bibitem#1\endcsname}%
\let\auto@bib@innerbib\@empty
\bibitem [{\citenamefont {Terhal}\ \emph {et~al.}(2020)\citenamefont {Terhal},
  \citenamefont {Conrad},\ and\ \citenamefont {Vuillot}}]{Terhal_2020}%
  \BibitemOpen
  \bibfield  {author} {\bibinfo {author} {\bibfnamefont {B.~M.}\ \bibnamefont
  {Terhal}}, \bibinfo {author} {\bibfnamefont {J.}~\bibnamefont {Conrad}},\
  and\ \bibinfo {author} {\bibfnamefont {C.}~\bibnamefont {Vuillot}},\ }\href
  {https://doi.org/10.1088/2058-9565/ab98a5} {\bibfield  {journal} {\bibinfo
  {journal} {Quantum Science and Technology}\ }\textbf {\bibinfo {volume}
  {5}},\ \bibinfo {pages} {043001} (\bibinfo {year} {2020})}\BibitemShut
  {NoStop}%
\bibitem [{\citenamefont {Fukui}\ \emph {et~al.}(2018)\citenamefont {Fukui},
  \citenamefont {Tomita}, \citenamefont {Okamoto},\ and\ \citenamefont
  {Fujii}}]{PhysRevX.8.021054}%
  \BibitemOpen
  \bibfield  {author} {\bibinfo {author} {\bibfnamefont {K.}~\bibnamefont
  {Fukui}}, \bibinfo {author} {\bibfnamefont {A.}~\bibnamefont {Tomita}},
  \bibinfo {author} {\bibfnamefont {A.}~\bibnamefont {Okamoto}},\ and\ \bibinfo
  {author} {\bibfnamefont {K.}~\bibnamefont {Fujii}},\ }\href
  {https://doi.org/10.1103/PhysRevX.8.021054} {\bibfield  {journal} {\bibinfo
  {journal} {Phys. Rev. X}\ }\textbf {\bibinfo {volume} {8}},\ \bibinfo {pages}
  {021054} (\bibinfo {year} {2018})}\BibitemShut {NoStop}%
\bibitem [{\citenamefont {Grimsmo}\ and\ \citenamefont
  {Puri}(2021)}]{PRXQuantum.2.020101}%
  \BibitemOpen
  \bibfield  {author} {\bibinfo {author} {\bibfnamefont {A.~L.}\ \bibnamefont
  {Grimsmo}}\ and\ \bibinfo {author} {\bibfnamefont {S.}~\bibnamefont {Puri}},\
  }\href {https://doi.org/10.1103/PRXQuantum.2.020101} {\bibfield  {journal}
  {\bibinfo  {journal} {PRX Quantum}\ }\textbf {\bibinfo {volume} {2}},\
  \bibinfo {pages} {020101} (\bibinfo {year} {2021})}\BibitemShut {NoStop}%
\bibitem [{\citenamefont {Darmawan}\ \emph {et~al.}(2021)\citenamefont
  {Darmawan}, \citenamefont {Brown}, \citenamefont {Grimsmo}, \citenamefont
  {Tuckett},\ and\ \citenamefont {Puri}}]{PRXQuantum.2.030345}%
  \BibitemOpen
  \bibfield  {author} {\bibinfo {author} {\bibfnamefont {A.~S.}\ \bibnamefont
  {Darmawan}}, \bibinfo {author} {\bibfnamefont {B.~J.}\ \bibnamefont {Brown}},
  \bibinfo {author} {\bibfnamefont {A.~L.}\ \bibnamefont {Grimsmo}}, \bibinfo
  {author} {\bibfnamefont {D.~K.}\ \bibnamefont {Tuckett}},\ and\ \bibinfo
  {author} {\bibfnamefont {S.}~\bibnamefont {Puri}},\ }\href
  {https://doi.org/10.1103/PRXQuantum.2.030345} {\bibfield  {journal} {\bibinfo
   {journal} {PRX Quantum}\ }\textbf {\bibinfo {volume} {2}},\ \bibinfo {pages}
  {030345} (\bibinfo {year} {2021})}\BibitemShut {NoStop}%
\bibitem [{\citenamefont {Tzitrin}\ \emph {et~al.}(2021)\citenamefont
  {Tzitrin}, \citenamefont {Matsuura}, \citenamefont {Alexander}, \citenamefont
  {Dauphinais}, \citenamefont {Bourassa}, \citenamefont {Sabapathy},
  \citenamefont {Menicucci},\ and\ \citenamefont
  {Dhand}}]{PRXQuantum.2.040353}%
  \BibitemOpen
  \bibfield  {author} {\bibinfo {author} {\bibfnamefont {I.}~\bibnamefont
  {Tzitrin}}, \bibinfo {author} {\bibfnamefont {T.}~\bibnamefont {Matsuura}},
  \bibinfo {author} {\bibfnamefont {R.~N.}\ \bibnamefont {Alexander}}, \bibinfo
  {author} {\bibfnamefont {G.}~\bibnamefont {Dauphinais}}, \bibinfo {author}
  {\bibfnamefont {J.~E.}\ \bibnamefont {Bourassa}}, \bibinfo {author}
  {\bibfnamefont {K.~K.}\ \bibnamefont {Sabapathy}}, \bibinfo {author}
  {\bibfnamefont {N.~C.}\ \bibnamefont {Menicucci}},\ and\ \bibinfo {author}
  {\bibfnamefont {I.}~\bibnamefont {Dhand}},\ }\href
  {https://doi.org/10.1103/PRXQuantum.2.040353} {\bibfield  {journal} {\bibinfo
   {journal} {PRX Quantum}\ }\textbf {\bibinfo {volume} {2}},\ \bibinfo {pages}
  {040353} (\bibinfo {year} {2021})}\BibitemShut {NoStop}%
\bibitem [{\citenamefont {Chamberland}\ \emph {et~al.}(2022)\citenamefont
  {Chamberland}, \citenamefont {Noh}, \citenamefont {Arrangoiz-Arriola},
  \citenamefont {Campbell}, \citenamefont {Hann}, \citenamefont {Iverson},
  \citenamefont {Putterman}, \citenamefont {Bohdanowicz}, \citenamefont
  {Flammia}, \citenamefont {Keller}, \citenamefont {Refael}, \citenamefont
  {Preskill}, \citenamefont {Jiang}, \citenamefont {Safavi-Naeini},
  \citenamefont {Painter},\ and\ \citenamefont
  {Brand\~ao}}]{PRXQuantum.3.010329}%
  \BibitemOpen
  \bibfield  {author} {\bibinfo {author} {\bibfnamefont {C.}~\bibnamefont
  {Chamberland}}, \bibinfo {author} {\bibfnamefont {K.}~\bibnamefont {Noh}},
  \bibinfo {author} {\bibfnamefont {P.}~\bibnamefont {Arrangoiz-Arriola}},
  \bibinfo {author} {\bibfnamefont {E.~T.}\ \bibnamefont {Campbell}}, \bibinfo
  {author} {\bibfnamefont {C.~T.}\ \bibnamefont {Hann}}, \bibinfo {author}
  {\bibfnamefont {J.}~\bibnamefont {Iverson}}, \bibinfo {author} {\bibfnamefont
  {H.}~\bibnamefont {Putterman}}, \bibinfo {author} {\bibfnamefont {T.~C.}\
  \bibnamefont {Bohdanowicz}}, \bibinfo {author} {\bibfnamefont {S.~T.}\
  \bibnamefont {Flammia}}, \bibinfo {author} {\bibfnamefont {A.}~\bibnamefont
  {Keller}}, \bibinfo {author} {\bibfnamefont {G.}~\bibnamefont {Refael}},
  \bibinfo {author} {\bibfnamefont {J.}~\bibnamefont {Preskill}}, \bibinfo
  {author} {\bibfnamefont {L.}~\bibnamefont {Jiang}}, \bibinfo {author}
  {\bibfnamefont {A.~H.}\ \bibnamefont {Safavi-Naeini}}, \bibinfo {author}
  {\bibfnamefont {O.}~\bibnamefont {Painter}},\ and\ \bibinfo {author}
  {\bibfnamefont {F.~G.}\ \bibnamefont {Brand\~ao}},\ }\href
  {https://doi.org/10.1103/PRXQuantum.3.010329} {\bibfield  {journal} {\bibinfo
   {journal} {PRX Quantum}\ }\textbf {\bibinfo {volume} {3}},\ \bibinfo {pages}
  {010329} (\bibinfo {year} {2022})}\BibitemShut {NoStop}%
\bibitem [{\citenamefont {Baragiola}\ \emph {et~al.}(2019)\citenamefont
  {Baragiola}, \citenamefont {Pantaleoni}, \citenamefont {Alexander},
  \citenamefont {Karanjai},\ and\ \citenamefont
  {Menicucci}}]{PhysRevLett.123.200502}%
  \BibitemOpen
  \bibfield  {author} {\bibinfo {author} {\bibfnamefont {B.~Q.}\ \bibnamefont
  {Baragiola}}, \bibinfo {author} {\bibfnamefont {G.}~\bibnamefont
  {Pantaleoni}}, \bibinfo {author} {\bibfnamefont {R.~N.}\ \bibnamefont
  {Alexander}}, \bibinfo {author} {\bibfnamefont {A.}~\bibnamefont
  {Karanjai}},\ and\ \bibinfo {author} {\bibfnamefont {N.~C.}\ \bibnamefont
  {Menicucci}},\ }\href {https://doi.org/10.1103/PhysRevLett.123.200502}
  {\bibfield  {journal} {\bibinfo  {journal} {Phys. Rev. Lett.}\ }\textbf
  {\bibinfo {volume} {123}},\ \bibinfo {pages} {200502} (\bibinfo {year}
  {2019})}\BibitemShut {NoStop}%
\bibitem [{\citenamefont {Campagne-Ibarcq}\ \emph {et~al.}(2020)\citenamefont
  {Campagne-Ibarcq}, \citenamefont {Eickbusch}, \citenamefont {Touzard},
  \citenamefont {Zalys-Geller}, \citenamefont {Frattini}, \citenamefont
  {Sivak}, \citenamefont {Reinhold}, \citenamefont {Puri}, \citenamefont
  {Shankar}, \citenamefont {Schoelkopf}, \citenamefont {Frunzio}, \citenamefont
  {Mirrahimi},\ and\ \citenamefont {Devoret}}]{ExpErrorCorr1}%
  \BibitemOpen
  \bibfield  {author} {\bibinfo {author} {\bibfnamefont {P.}~\bibnamefont
  {Campagne-Ibarcq}}, \bibinfo {author} {\bibfnamefont {A.}~\bibnamefont
  {Eickbusch}}, \bibinfo {author} {\bibfnamefont {S.}~\bibnamefont {Touzard}},
  \bibinfo {author} {\bibfnamefont {E.}~\bibnamefont {Zalys-Geller}}, \bibinfo
  {author} {\bibfnamefont {N.~E.}\ \bibnamefont {Frattini}}, \bibinfo {author}
  {\bibfnamefont {V.~V.}\ \bibnamefont {Sivak}}, \bibinfo {author}
  {\bibfnamefont {P.}~\bibnamefont {Reinhold}}, \bibinfo {author}
  {\bibfnamefont {S.}~\bibnamefont {Puri}}, \bibinfo {author} {\bibfnamefont
  {S.}~\bibnamefont {Shankar}}, \bibinfo {author} {\bibfnamefont {R.~J.}\
  \bibnamefont {Schoelkopf}}, \bibinfo {author} {\bibfnamefont
  {L.}~\bibnamefont {Frunzio}}, \bibinfo {author} {\bibfnamefont
  {M.}~\bibnamefont {Mirrahimi}},\ and\ \bibinfo {author} {\bibfnamefont
  {M.~H.}\ \bibnamefont {Devoret}},\ }\href
  {https://doi.org/10.1038/s41586-020-2603-3} {\bibfield  {journal} {\bibinfo
  {journal} {Nature}\ }\textbf {\bibinfo {volume} {584}},\ \bibinfo {pages}
  {368} (\bibinfo {year} {2020})}\BibitemShut {NoStop}%
\bibitem [{\citenamefont {de~Neeve}\ \emph {et~al.}(2022)\citenamefont
  {de~Neeve}, \citenamefont {Nguyen}, \citenamefont {Behrle},\ and\
  \citenamefont {Home}}]{ExpErrorCorr2}%
  \BibitemOpen
  \bibfield  {author} {\bibinfo {author} {\bibfnamefont {B.}~\bibnamefont
  {de~Neeve}}, \bibinfo {author} {\bibfnamefont {T.-L.}\ \bibnamefont
  {Nguyen}}, \bibinfo {author} {\bibfnamefont {T.}~\bibnamefont {Behrle}},\
  and\ \bibinfo {author} {\bibfnamefont {J.~P.}\ \bibnamefont {Home}},\ }\href
  {https://doi.org/10.1038/s41567-021-01487-7} {\bibfield  {journal} {\bibinfo
  {journal} {Nature Physics}\ }\textbf {\bibinfo {volume} {18}},\ \bibinfo
  {pages} {296} (\bibinfo {year} {2022})}\BibitemShut {NoStop}%
\bibitem [{\citenamefont {Hu}\ \emph {et~al.}(2019)\citenamefont {Hu},
  \citenamefont {Ma}, \citenamefont {Cai}, \citenamefont {Mu}, \citenamefont
  {Xu}, \citenamefont {Wang}, \citenamefont {Wu}, \citenamefont {Wang},
  \citenamefont {Song}, \citenamefont {Zou}, \citenamefont {Girvin},
  \citenamefont {Duan},\ and\ \citenamefont {Sun}}]{ExpErrorCorr3}%
  \BibitemOpen
  \bibfield  {author} {\bibinfo {author} {\bibfnamefont {L.}~\bibnamefont
  {Hu}}, \bibinfo {author} {\bibfnamefont {Y.}~\bibnamefont {Ma}}, \bibinfo
  {author} {\bibfnamefont {W.}~\bibnamefont {Cai}}, \bibinfo {author}
  {\bibfnamefont {X.}~\bibnamefont {Mu}}, \bibinfo {author} {\bibfnamefont
  {Y.}~\bibnamefont {Xu}}, \bibinfo {author} {\bibfnamefont {W.}~\bibnamefont
  {Wang}}, \bibinfo {author} {\bibfnamefont {Y.}~\bibnamefont {Wu}}, \bibinfo
  {author} {\bibfnamefont {H.}~\bibnamefont {Wang}}, \bibinfo {author}
  {\bibfnamefont {Y.~P.}\ \bibnamefont {Song}}, \bibinfo {author}
  {\bibfnamefont {C.~L.}\ \bibnamefont {Zou}}, \bibinfo {author} {\bibfnamefont
  {S.~M.}\ \bibnamefont {Girvin}}, \bibinfo {author} {\bibfnamefont {L.-M.}\
  \bibnamefont {Duan}},\ and\ \bibinfo {author} {\bibfnamefont
  {L.}~\bibnamefont {Sun}},\ }\href {https://doi.org/10.1038/s41567-018-0414-3}
  {\bibfield  {journal} {\bibinfo  {journal} {Nature Physics}\ }\textbf
  {\bibinfo {volume} {15}},\ \bibinfo {pages} {503} (\bibinfo {year}
  {2019})}\BibitemShut {NoStop}%
\bibitem [{\citenamefont {Lescanne}\ \emph {et~al.}(2020)\citenamefont
  {Lescanne}, \citenamefont {Villiers}, \citenamefont {Peronnin}, \citenamefont
  {Sarlette}, \citenamefont {Delbecq}, \citenamefont {Huard}, \citenamefont
  {Kontos}, \citenamefont {Mirrahimi},\ and\ \citenamefont
  {Leghtas}}]{ExpErrorCorr4}%
  \BibitemOpen
  \bibfield  {author} {\bibinfo {author} {\bibfnamefont {R.}~\bibnamefont
  {Lescanne}}, \bibinfo {author} {\bibfnamefont {M.}~\bibnamefont {Villiers}},
  \bibinfo {author} {\bibfnamefont {T.}~\bibnamefont {Peronnin}}, \bibinfo
  {author} {\bibfnamefont {A.}~\bibnamefont {Sarlette}}, \bibinfo {author}
  {\bibfnamefont {M.}~\bibnamefont {Delbecq}}, \bibinfo {author} {\bibfnamefont
  {B.}~\bibnamefont {Huard}}, \bibinfo {author} {\bibfnamefont
  {T.}~\bibnamefont {Kontos}}, \bibinfo {author} {\bibfnamefont
  {M.}~\bibnamefont {Mirrahimi}},\ and\ \bibinfo {author} {\bibfnamefont
  {Z.}~\bibnamefont {Leghtas}},\ }\href
  {https://doi.org/10.1038/s41567-020-0824-x} {\bibfield  {journal} {\bibinfo
  {journal} {Nature Physics}\ }\textbf {\bibinfo {volume} {16}},\ \bibinfo
  {pages} {509} (\bibinfo {year} {2020})}\BibitemShut {NoStop}%
\bibitem [{\citenamefont {Hamilton}\ \emph {et~al.}(2017)\citenamefont
  {Hamilton}, \citenamefont {Kruse}, \citenamefont {Sansoni}, \citenamefont
  {Barkhofen}, \citenamefont {Silberhorn},\ and\ \citenamefont
  {Jex}}]{PhysRevLett.119.170501}%
  \BibitemOpen
  \bibfield  {author} {\bibinfo {author} {\bibfnamefont {C.~S.}\ \bibnamefont
  {Hamilton}}, \bibinfo {author} {\bibfnamefont {R.}~\bibnamefont {Kruse}},
  \bibinfo {author} {\bibfnamefont {L.}~\bibnamefont {Sansoni}}, \bibinfo
  {author} {\bibfnamefont {S.}~\bibnamefont {Barkhofen}}, \bibinfo {author}
  {\bibfnamefont {C.}~\bibnamefont {Silberhorn}},\ and\ \bibinfo {author}
  {\bibfnamefont {I.}~\bibnamefont {Jex}},\ }\href
  {https://doi.org/10.1103/PhysRevLett.119.170501} {\bibfield  {journal}
  {\bibinfo  {journal} {Phys. Rev. Lett.}\ }\textbf {\bibinfo {volume} {119}},\
  \bibinfo {pages} {170501} (\bibinfo {year} {2017})}\BibitemShut {NoStop}%
\bibitem [{\citenamefont {Quesada}\ \emph {et~al.}(2018)\citenamefont
  {Quesada}, \citenamefont {Arrazola},\ and\ \citenamefont
  {Killoran}}]{PhysRevA.98.062322}%
  \BibitemOpen
  \bibfield  {author} {\bibinfo {author} {\bibfnamefont {N.}~\bibnamefont
  {Quesada}}, \bibinfo {author} {\bibfnamefont {J.~M.}\ \bibnamefont
  {Arrazola}},\ and\ \bibinfo {author} {\bibfnamefont {N.}~\bibnamefont
  {Killoran}},\ }\href {https://doi.org/10.1103/PhysRevA.98.062322} {\bibfield
  {journal} {\bibinfo  {journal} {Phys. Rev. A}\ }\textbf {\bibinfo {volume}
  {98}},\ \bibinfo {pages} {062322} (\bibinfo {year} {2018})}\BibitemShut
  {NoStop}%
\bibitem [{\citenamefont {Quesada}\ \emph {et~al.}(2022)\citenamefont
  {Quesada}, \citenamefont {Chadwick}, \citenamefont {Bell}, \citenamefont
  {Arrazola}, \citenamefont {Vincent}, \citenamefont {Qi},\ and\ \citenamefont
  {Garc\'{\i}a\ensuremath{-}Patr\'on}}]{PRXQuantum.3.010306}%
  \BibitemOpen
  \bibfield  {author} {\bibinfo {author} {\bibfnamefont {N.}~\bibnamefont
  {Quesada}}, \bibinfo {author} {\bibfnamefont {R.~S.}\ \bibnamefont
  {Chadwick}}, \bibinfo {author} {\bibfnamefont {B.~A.}\ \bibnamefont {Bell}},
  \bibinfo {author} {\bibfnamefont {J.~M.}\ \bibnamefont {Arrazola}}, \bibinfo
  {author} {\bibfnamefont {T.}~\bibnamefont {Vincent}}, \bibinfo {author}
  {\bibfnamefont {H.}~\bibnamefont {Qi}},\ and\ \bibinfo {author}
  {\bibfnamefont {R.}~\bibnamefont {Garc\'{\i}a\ensuremath{-}Patr\'on}},\
  }\href {https://doi.org/10.1103/PRXQuantum.3.010306} {\bibfield  {journal}
  {\bibinfo  {journal} {PRX Quantum}\ }\textbf {\bibinfo {volume} {3}},\
  \bibinfo {pages} {010306} (\bibinfo {year} {2022})}\BibitemShut {NoStop}%
\bibitem [{\citenamefont {Bulmer}\ \emph {et~al.}(2022)\citenamefont {Bulmer},
  \citenamefont {Bell}, \citenamefont {Chadwick}, \citenamefont {Jones},
  \citenamefont {Moise}, \citenamefont {Rigazzi}, \citenamefont {Thorbecke},
  \citenamefont {Haus}, \citenamefont {Vaerenbergh}, \citenamefont {Patel},
  \citenamefont {Walmsley},\ and\ \citenamefont
  {Laing}}]{doi:10.1126/sciadv.abl9236}%
  \BibitemOpen
  \bibfield  {author} {\bibinfo {author} {\bibfnamefont {J.~F.~F.}\
  \bibnamefont {Bulmer}}, \bibinfo {author} {\bibfnamefont {B.~A.}\
  \bibnamefont {Bell}}, \bibinfo {author} {\bibfnamefont {R.~S.}\ \bibnamefont
  {Chadwick}}, \bibinfo {author} {\bibfnamefont {A.~E.}\ \bibnamefont {Jones}},
  \bibinfo {author} {\bibfnamefont {D.}~\bibnamefont {Moise}}, \bibinfo
  {author} {\bibfnamefont {A.}~\bibnamefont {Rigazzi}}, \bibinfo {author}
  {\bibfnamefont {J.}~\bibnamefont {Thorbecke}}, \bibinfo {author}
  {\bibfnamefont {U.-U.}\ \bibnamefont {Haus}}, \bibinfo {author}
  {\bibfnamefont {T.~V.}\ \bibnamefont {Vaerenbergh}}, \bibinfo {author}
  {\bibfnamefont {R.~B.}\ \bibnamefont {Patel}}, \bibinfo {author}
  {\bibfnamefont {I.~A.}\ \bibnamefont {Walmsley}},\ and\ \bibinfo {author}
  {\bibfnamefont {A.}~\bibnamefont {Laing}},\ }\href
  {https://doi.org/10.1126/sciadv.abl9236} {\bibfield  {journal} {\bibinfo
  {journal} {Science Advances}\ }\textbf {\bibinfo {volume} {8}},\ \bibinfo
  {pages} {eabl9236} (\bibinfo {year} {2022})}\BibitemShut {NoStop}%
\bibitem [{\citenamefont {Zhong}\ \emph {et~al.}(2020)\citenamefont {Zhong},
  \citenamefont {Wang}, \citenamefont {Deng}, \citenamefont {Chen},
  \citenamefont {Peng}, \citenamefont {Luo}, \citenamefont {Qin}, \citenamefont
  {Wu}, \citenamefont {Ding}, \citenamefont {Hu}, \citenamefont {Hu},
  \citenamefont {Yang}, \citenamefont {Zhang}, \citenamefont {Li},
  \citenamefont {Li}, \citenamefont {Jiang}, \citenamefont {Gan}, \citenamefont
  {Yang}, \citenamefont {You}, \citenamefont {Wang}, \citenamefont {Li},
  \citenamefont {Liu}, \citenamefont {Lu},\ and\ \citenamefont
  {Pan}}]{doi:10.1126/science.abe8770}%
  \BibitemOpen
  \bibfield  {author} {\bibinfo {author} {\bibfnamefont {H.-S.}\ \bibnamefont
  {Zhong}}, \bibinfo {author} {\bibfnamefont {H.}~\bibnamefont {Wang}},
  \bibinfo {author} {\bibfnamefont {Y.-H.}\ \bibnamefont {Deng}}, \bibinfo
  {author} {\bibfnamefont {M.-C.}\ \bibnamefont {Chen}}, \bibinfo {author}
  {\bibfnamefont {L.-C.}\ \bibnamefont {Peng}}, \bibinfo {author}
  {\bibfnamefont {Y.-H.}\ \bibnamefont {Luo}}, \bibinfo {author} {\bibfnamefont
  {J.}~\bibnamefont {Qin}}, \bibinfo {author} {\bibfnamefont {D.}~\bibnamefont
  {Wu}}, \bibinfo {author} {\bibfnamefont {X.}~\bibnamefont {Ding}}, \bibinfo
  {author} {\bibfnamefont {Y.}~\bibnamefont {Hu}}, \bibinfo {author}
  {\bibfnamefont {P.}~\bibnamefont {Hu}}, \bibinfo {author} {\bibfnamefont
  {X.-Y.}\ \bibnamefont {Yang}}, \bibinfo {author} {\bibfnamefont {W.-J.}\
  \bibnamefont {Zhang}}, \bibinfo {author} {\bibfnamefont {H.}~\bibnamefont
  {Li}}, \bibinfo {author} {\bibfnamefont {Y.}~\bibnamefont {Li}}, \bibinfo
  {author} {\bibfnamefont {X.}~\bibnamefont {Jiang}}, \bibinfo {author}
  {\bibfnamefont {L.}~\bibnamefont {Gan}}, \bibinfo {author} {\bibfnamefont
  {G.}~\bibnamefont {Yang}}, \bibinfo {author} {\bibfnamefont {L.}~\bibnamefont
  {You}}, \bibinfo {author} {\bibfnamefont {Z.}~\bibnamefont {Wang}}, \bibinfo
  {author} {\bibfnamefont {L.}~\bibnamefont {Li}}, \bibinfo {author}
  {\bibfnamefont {N.-L.}\ \bibnamefont {Liu}}, \bibinfo {author} {\bibfnamefont
  {C.-Y.}\ \bibnamefont {Lu}},\ and\ \bibinfo {author} {\bibfnamefont {J.-W.}\
  \bibnamefont {Pan}},\ }\href {https://doi.org/10.1126/science.abe8770}
  {\bibfield  {journal} {\bibinfo  {journal} {Science}\ }\textbf {\bibinfo
  {volume} {370}},\ \bibinfo {pages} {1460} (\bibinfo {year}
  {2020})}\BibitemShut {NoStop}%
\bibitem [{\citenamefont {Zhong}\ \emph {et~al.}(2021)\citenamefont {Zhong},
  \citenamefont {Deng}, \citenamefont {Qin}, \citenamefont {Wang},
  \citenamefont {Chen}, \citenamefont {Peng}, \citenamefont {Luo},
  \citenamefont {Wu}, \citenamefont {Gong}, \citenamefont {Su}, \citenamefont
  {Hu}, \citenamefont {Hu}, \citenamefont {Yang}, \citenamefont {Zhang},
  \citenamefont {Li}, \citenamefont {Li}, \citenamefont {Jiang}, \citenamefont
  {Gan}, \citenamefont {Yang}, \citenamefont {You}, \citenamefont {Wang},
  \citenamefont {Li}, \citenamefont {Liu}, \citenamefont {Renema},
  \citenamefont {Lu},\ and\ \citenamefont {Pan}}]{PhysRevLett.127.180502}%
  \BibitemOpen
  \bibfield  {author} {\bibinfo {author} {\bibfnamefont {H.-S.}\ \bibnamefont
  {Zhong}}, \bibinfo {author} {\bibfnamefont {Y.-H.}\ \bibnamefont {Deng}},
  \bibinfo {author} {\bibfnamefont {J.}~\bibnamefont {Qin}}, \bibinfo {author}
  {\bibfnamefont {H.}~\bibnamefont {Wang}}, \bibinfo {author} {\bibfnamefont
  {M.-C.}\ \bibnamefont {Chen}}, \bibinfo {author} {\bibfnamefont {L.-C.}\
  \bibnamefont {Peng}}, \bibinfo {author} {\bibfnamefont {Y.-H.}\ \bibnamefont
  {Luo}}, \bibinfo {author} {\bibfnamefont {D.}~\bibnamefont {Wu}}, \bibinfo
  {author} {\bibfnamefont {S.-Q.}\ \bibnamefont {Gong}}, \bibinfo {author}
  {\bibfnamefont {H.}~\bibnamefont {Su}}, \bibinfo {author} {\bibfnamefont
  {Y.}~\bibnamefont {Hu}}, \bibinfo {author} {\bibfnamefont {P.}~\bibnamefont
  {Hu}}, \bibinfo {author} {\bibfnamefont {X.-Y.}\ \bibnamefont {Yang}},
  \bibinfo {author} {\bibfnamefont {W.-J.}\ \bibnamefont {Zhang}}, \bibinfo
  {author} {\bibfnamefont {H.}~\bibnamefont {Li}}, \bibinfo {author}
  {\bibfnamefont {Y.}~\bibnamefont {Li}}, \bibinfo {author} {\bibfnamefont
  {X.}~\bibnamefont {Jiang}}, \bibinfo {author} {\bibfnamefont
  {L.}~\bibnamefont {Gan}}, \bibinfo {author} {\bibfnamefont {G.}~\bibnamefont
  {Yang}}, \bibinfo {author} {\bibfnamefont {L.}~\bibnamefont {You}}, \bibinfo
  {author} {\bibfnamefont {Z.}~\bibnamefont {Wang}}, \bibinfo {author}
  {\bibfnamefont {L.}~\bibnamefont {Li}}, \bibinfo {author} {\bibfnamefont
  {N.-L.}\ \bibnamefont {Liu}}, \bibinfo {author} {\bibfnamefont {J.~J.}\
  \bibnamefont {Renema}}, \bibinfo {author} {\bibfnamefont {C.-Y.}\
  \bibnamefont {Lu}},\ and\ \bibinfo {author} {\bibfnamefont {J.-W.}\
  \bibnamefont {Pan}},\ }\href {https://doi.org/10.1103/PhysRevLett.127.180502}
  {\bibfield  {journal} {\bibinfo  {journal} {Phys. Rev. Lett.}\ }\textbf
  {\bibinfo {volume} {127}},\ \bibinfo {pages} {180502} (\bibinfo {year}
  {2021})}\BibitemShut {NoStop}%
\bibitem [{\citenamefont {Madsen}\ \emph {et~al.}(2022)\citenamefont {Madsen},
  \citenamefont {Laudenbach}, \citenamefont {Askarani}, \citenamefont
  {Rortais}, \citenamefont {Vincent}, \citenamefont {Bulmer}, \citenamefont
  {Miatto}, \citenamefont {Neuhaus}, \citenamefont {Helt}, \citenamefont
  {Collins}, \citenamefont {Lita}, \citenamefont {Gerrits}, \citenamefont
  {Nam}, \citenamefont {Vaidya}, \citenamefont {Menotti}, \citenamefont
  {Dhand}, \citenamefont {Vernon}, \citenamefont {Quesada},\ and\ \citenamefont
  {Lavoie}}]{Borealis}%
  \BibitemOpen
  \bibfield  {author} {\bibinfo {author} {\bibfnamefont {L.~S.}\ \bibnamefont
  {Madsen}}, \bibinfo {author} {\bibfnamefont {F.}~\bibnamefont {Laudenbach}},
  \bibinfo {author} {\bibfnamefont {M.~F.}\ \bibnamefont {Askarani}}, \bibinfo
  {author} {\bibfnamefont {F.}~\bibnamefont {Rortais}}, \bibinfo {author}
  {\bibfnamefont {T.}~\bibnamefont {Vincent}}, \bibinfo {author} {\bibfnamefont
  {J.~F.~F.}\ \bibnamefont {Bulmer}}, \bibinfo {author} {\bibfnamefont {F.~M.}\
  \bibnamefont {Miatto}}, \bibinfo {author} {\bibfnamefont {L.}~\bibnamefont
  {Neuhaus}}, \bibinfo {author} {\bibfnamefont {L.~G.}\ \bibnamefont {Helt}},
  \bibinfo {author} {\bibfnamefont {M.~J.}\ \bibnamefont {Collins}}, \bibinfo
  {author} {\bibfnamefont {A.~E.}\ \bibnamefont {Lita}}, \bibinfo {author}
  {\bibfnamefont {T.}~\bibnamefont {Gerrits}}, \bibinfo {author} {\bibfnamefont
  {S.~W.}\ \bibnamefont {Nam}}, \bibinfo {author} {\bibfnamefont {V.~D.}\
  \bibnamefont {Vaidya}}, \bibinfo {author} {\bibfnamefont {M.}~\bibnamefont
  {Menotti}}, \bibinfo {author} {\bibfnamefont {I.}~\bibnamefont {Dhand}},
  \bibinfo {author} {\bibfnamefont {Z.}~\bibnamefont {Vernon}}, \bibinfo
  {author} {\bibfnamefont {N.}~\bibnamefont {Quesada}},\ and\ \bibinfo {author}
  {\bibfnamefont {J.}~\bibnamefont {Lavoie}},\ }\href
  {https://doi.org/10.1038/s41586-022-04725-x} {\bibfield  {journal} {\bibinfo
  {journal} {Nature}\ }\textbf {\bibinfo {volume} {606}},\ \bibinfo {pages}
  {75} (\bibinfo {year} {2022})}\BibitemShut {NoStop}%
\bibitem [{\citenamefont {Weedbrook}\ \emph {et~al.}(2012)\citenamefont
  {Weedbrook}, \citenamefont {Pirandola}, \citenamefont {Garc\'{\i}a-Patr\'on},
  \citenamefont {Cerf}, \citenamefont {Ralph}, \citenamefont {Shapiro},\ and\
  \citenamefont {Lloyd}}]{RevModPhys.84.621}%
  \BibitemOpen
  \bibfield  {author} {\bibinfo {author} {\bibfnamefont {C.}~\bibnamefont
  {Weedbrook}}, \bibinfo {author} {\bibfnamefont {S.}~\bibnamefont
  {Pirandola}}, \bibinfo {author} {\bibfnamefont {R.}~\bibnamefont
  {Garc\'{\i}a-Patr\'on}}, \bibinfo {author} {\bibfnamefont {N.~J.}\
  \bibnamefont {Cerf}}, \bibinfo {author} {\bibfnamefont {T.~C.}\ \bibnamefont
  {Ralph}}, \bibinfo {author} {\bibfnamefont {J.~H.}\ \bibnamefont {Shapiro}},\
  and\ \bibinfo {author} {\bibfnamefont {S.}~\bibnamefont {Lloyd}},\ }\href
  {https://doi.org/10.1103/RevModPhys.84.621} {\bibfield  {journal} {\bibinfo
  {journal} {Rev. Mod. Phys.}\ }\textbf {\bibinfo {volume} {84}},\ \bibinfo
  {pages} {621} (\bibinfo {year} {2012})}\BibitemShut {NoStop}%
\bibitem [{\citenamefont {Fabre}\ and\ \citenamefont
  {Treps}(2020)}]{RevModPhys.92.035005}%
  \BibitemOpen
  \bibfield  {author} {\bibinfo {author} {\bibfnamefont {C.}~\bibnamefont
  {Fabre}}\ and\ \bibinfo {author} {\bibfnamefont {N.}~\bibnamefont {Treps}},\
  }\href {https://doi.org/10.1103/RevModPhys.92.035005} {\bibfield  {journal}
  {\bibinfo  {journal} {Rev. Mod. Phys.}\ }\textbf {\bibinfo {volume} {92}},\
  \bibinfo {pages} {035005} (\bibinfo {year} {2020})}\BibitemShut {NoStop}%
\bibitem [{\citenamefont {Walschaers}(2021)}]{PRXQuantum.2.030204}%
  \BibitemOpen
  \bibfield  {author} {\bibinfo {author} {\bibfnamefont {M.}~\bibnamefont
  {Walschaers}},\ }\href {https://doi.org/10.1103/PRXQuantum.2.030204}
  {\bibfield  {journal} {\bibinfo  {journal} {PRX Quantum}\ }\textbf {\bibinfo
  {volume} {2}},\ \bibinfo {pages} {030204} (\bibinfo {year}
  {2021})}\BibitemShut {NoStop}%
\bibitem [{\citenamefont {Su}\ \emph {et~al.}(2012)\citenamefont {Su},
  \citenamefont {Zhao}, \citenamefont {Hao}, \citenamefont {Jia}, \citenamefont
  {Xie},\ and\ \citenamefont {Peng}}]{Su:12}%
  \BibitemOpen
  \bibfield  {author} {\bibinfo {author} {\bibfnamefont {X.}~\bibnamefont
  {Su}}, \bibinfo {author} {\bibfnamefont {Y.}~\bibnamefont {Zhao}}, \bibinfo
  {author} {\bibfnamefont {S.}~\bibnamefont {Hao}}, \bibinfo {author}
  {\bibfnamefont {X.}~\bibnamefont {Jia}}, \bibinfo {author} {\bibfnamefont
  {C.}~\bibnamefont {Xie}},\ and\ \bibinfo {author} {\bibfnamefont
  {K.}~\bibnamefont {Peng}},\ }\href {https://doi.org/10.1364/OL.37.005178}
  {\bibfield  {journal} {\bibinfo  {journal} {Opt. Lett.}\ }\textbf {\bibinfo
  {volume} {37}},\ \bibinfo {pages} {5178} (\bibinfo {year}
  {2012})}\BibitemShut {NoStop}%
\bibitem [{\citenamefont {Cai}\ \emph {et~al.}(2017)\citenamefont {Cai},
  \citenamefont {Roslund}, \citenamefont {Ferrini}, \citenamefont {Arzani},
  \citenamefont {Xu}, \citenamefont {Fabre},\ and\ \citenamefont
  {Treps}}]{cai-2017}%
  \BibitemOpen
  \bibfield  {author} {\bibinfo {author} {\bibfnamefont {Y.}~\bibnamefont
  {Cai}}, \bibinfo {author} {\bibfnamefont {J.}~\bibnamefont {Roslund}},
  \bibinfo {author} {\bibfnamefont {G.}~\bibnamefont {Ferrini}}, \bibinfo
  {author} {\bibfnamefont {F.}~\bibnamefont {Arzani}}, \bibinfo {author}
  {\bibfnamefont {X.}~\bibnamefont {Xu}}, \bibinfo {author} {\bibfnamefont
  {C.}~\bibnamefont {Fabre}},\ and\ \bibinfo {author} {\bibfnamefont
  {N.}~\bibnamefont {Treps}},\ }\href {http://dx.doi.org/10.1038/ncomms15645}
  {\bibfield  {journal} {\bibinfo  {journal} {Nat. Commun.}\ }\textbf {\bibinfo
  {volume} {8}},\ \bibinfo {pages} {15645} (\bibinfo {year}
  {2017})}\BibitemShut {NoStop}%
\bibitem [{\citenamefont {Asavanant}\ \emph {et~al.}(2019)\citenamefont
  {Asavanant}, \citenamefont {Shiozawa}, \citenamefont {Yokoyama},
  \citenamefont {Charoensombutamon}, \citenamefont {Emura}, \citenamefont
  {Alexander}, \citenamefont {Takeda}, \citenamefont {Yoshikawa}, \citenamefont
  {Menicucci}, \citenamefont {Yonezawa},\ and\ \citenamefont
  {Furusawa}}]{Asavanant:2019aa}%
  \BibitemOpen
  \bibfield  {author} {\bibinfo {author} {\bibfnamefont {W.}~\bibnamefont
  {Asavanant}}, \bibinfo {author} {\bibfnamefont {Y.}~\bibnamefont {Shiozawa}},
  \bibinfo {author} {\bibfnamefont {S.}~\bibnamefont {Yokoyama}}, \bibinfo
  {author} {\bibfnamefont {B.}~\bibnamefont {Charoensombutamon}}, \bibinfo
  {author} {\bibfnamefont {H.}~\bibnamefont {Emura}}, \bibinfo {author}
  {\bibfnamefont {R.~N.}\ \bibnamefont {Alexander}}, \bibinfo {author}
  {\bibfnamefont {S.}~\bibnamefont {Takeda}}, \bibinfo {author} {\bibfnamefont
  {J.-i.}\ \bibnamefont {Yoshikawa}}, \bibinfo {author} {\bibfnamefont {N.~C.}\
  \bibnamefont {Menicucci}}, \bibinfo {author} {\bibfnamefont {H.}~\bibnamefont
  {Yonezawa}},\ and\ \bibinfo {author} {\bibfnamefont {A.}~\bibnamefont
  {Furusawa}},\ }\href {https://doi.org/10.1126/science.aay2645} {\bibfield
  {journal} {\bibinfo  {journal} {Science}\ }\textbf {\bibinfo {volume}
  {366}},\ \bibinfo {pages} {373} (\bibinfo {year} {2019})}\BibitemShut
  {NoStop}%
\bibitem [{\citenamefont {Larsen}\ \emph {et~al.}(2019)\citenamefont {Larsen},
  \citenamefont {Guo}, \citenamefont {Breum}, \citenamefont
  {Neergaard-Nielsen},\ and\ \citenamefont {Andersen}}]{Larsen:2019aa}%
  \BibitemOpen
  \bibfield  {author} {\bibinfo {author} {\bibfnamefont {M.~V.}\ \bibnamefont
  {Larsen}}, \bibinfo {author} {\bibfnamefont {X.}~\bibnamefont {Guo}},
  \bibinfo {author} {\bibfnamefont {C.~R.}\ \bibnamefont {Breum}}, \bibinfo
  {author} {\bibfnamefont {J.~S.}\ \bibnamefont {Neergaard-Nielsen}},\ and\
  \bibinfo {author} {\bibfnamefont {U.~L.}\ \bibnamefont {Andersen}},\ }\href
  {https://doi.org/10.1126/science.aay4354} {\bibfield  {journal} {\bibinfo
  {journal} {Science}\ }\textbf {\bibinfo {volume} {366}},\ \bibinfo {pages}
  {369} (\bibinfo {year} {2019})}\BibitemShut {NoStop}%
\bibitem [{\citenamefont {Mari}\ and\ \citenamefont
  {Eisert}(2012)}]{PhysRevLett.109.230503}%
  \BibitemOpen
  \bibfield  {author} {\bibinfo {author} {\bibfnamefont {A.}~\bibnamefont
  {Mari}}\ and\ \bibinfo {author} {\bibfnamefont {J.}~\bibnamefont {Eisert}},\
  }\href {https://doi.org/10.1103/PhysRevLett.109.230503} {\bibfield  {journal}
  {\bibinfo  {journal} {Phys. Rev. Lett.}\ }\textbf {\bibinfo {volume} {109}},\
  \bibinfo {pages} {230503} (\bibinfo {year} {2012})}\BibitemShut {NoStop}%
\bibitem [{\citenamefont {Veitch}\ \emph {et~al.}(2012)\citenamefont {Veitch},
  \citenamefont {Ferrie}, \citenamefont {Gross},\ and\ \citenamefont
  {Emerson}}]{Veitch_2012}%
  \BibitemOpen
  \bibfield  {author} {\bibinfo {author} {\bibfnamefont {V.}~\bibnamefont
  {Veitch}}, \bibinfo {author} {\bibfnamefont {C.}~\bibnamefont {Ferrie}},
  \bibinfo {author} {\bibfnamefont {D.}~\bibnamefont {Gross}},\ and\ \bibinfo
  {author} {\bibfnamefont {J.}~\bibnamefont {Emerson}},\ }\href
  {https://doi.org/10.1088/1367-2630/14/11/113011} {\bibfield  {journal}
  {\bibinfo  {journal} {New Journal of Physics}\ }\textbf {\bibinfo {volume}
  {14}},\ \bibinfo {pages} {113011} (\bibinfo {year} {2012})}\BibitemShut
  {NoStop}%
\bibitem [{\citenamefont {Chabaud}\ and\ \citenamefont
  {Walschaers}(2023)}]{CW2022}%
  \BibitemOpen
  \bibfield  {author} {\bibinfo {author} {\bibfnamefont {U.}~\bibnamefont
  {Chabaud}}\ and\ \bibinfo {author} {\bibfnamefont {M.}~\bibnamefont
  {Walschaers}},\ }\href {https://doi.org/10.1103/PhysRevLett.130.090602}
  {\bibfield  {journal} {\bibinfo  {journal} {Phys. Rev. Lett.}\ }\textbf
  {\bibinfo {volume} {130}},\ \bibinfo {pages} {090602} (\bibinfo {year}
  {2023})}\BibitemShut {NoStop}%
\bibitem [{\citenamefont {Walschaers}\ and\ \citenamefont
  {Treps}(2020)}]{PhysRevLett.124.150501}%
  \BibitemOpen
  \bibfield  {author} {\bibinfo {author} {\bibfnamefont {M.}~\bibnamefont
  {Walschaers}}\ and\ \bibinfo {author} {\bibfnamefont {N.}~\bibnamefont
  {Treps}},\ }\href {https://doi.org/10.1103/PhysRevLett.124.150501} {\bibfield
   {journal} {\bibinfo  {journal} {Phys. Rev. Lett.}\ }\textbf {\bibinfo
  {volume} {124}},\ \bibinfo {pages} {150501} (\bibinfo {year}
  {2020})}\BibitemShut {NoStop}%
\bibitem [{\citenamefont {Walschaers}\ \emph {et~al.}(2020)\citenamefont
  {Walschaers}, \citenamefont {Parigi},\ and\ \citenamefont
  {Treps}}]{PRXQuantum.1.020305}%
  \BibitemOpen
  \bibfield  {author} {\bibinfo {author} {\bibfnamefont {M.}~\bibnamefont
  {Walschaers}}, \bibinfo {author} {\bibfnamefont {V.}~\bibnamefont {Parigi}},\
  and\ \bibinfo {author} {\bibfnamefont {N.}~\bibnamefont {Treps}},\ }\href
  {https://doi.org/10.1103/PRXQuantum.1.020305} {\bibfield  {journal} {\bibinfo
   {journal} {PRX Quantum}\ }\textbf {\bibinfo {volume} {1}},\ \bibinfo {pages}
  {020305} (\bibinfo {year} {2020})}\BibitemShut {NoStop}%
\bibitem [{\citenamefont {Xiang}\ \emph {et~al.}(2022)\citenamefont {Xiang},
  \citenamefont {Liu}, \citenamefont {Guo}, \citenamefont {Gong}, \citenamefont
  {Treps}, \citenamefont {He},\ and\ \citenamefont {Walschaers}}]{Xiang2022}%
  \BibitemOpen
  \bibfield  {author} {\bibinfo {author} {\bibfnamefont {Y.}~\bibnamefont
  {Xiang}}, \bibinfo {author} {\bibfnamefont {S.}~\bibnamefont {Liu}}, \bibinfo
  {author} {\bibfnamefont {J.}~\bibnamefont {Guo}}, \bibinfo {author}
  {\bibfnamefont {Q.}~\bibnamefont {Gong}}, \bibinfo {author} {\bibfnamefont
  {N.}~\bibnamefont {Treps}}, \bibinfo {author} {\bibfnamefont
  {Q.}~\bibnamefont {He}},\ and\ \bibinfo {author} {\bibfnamefont
  {M.}~\bibnamefont {Walschaers}},\ }\href
  {https://doi.org/10.1038/s41534-022-00533-3} {\bibfield  {journal} {\bibinfo
  {journal} {npj Quantum Information}\ }\textbf {\bibinfo {volume} {8}},\
  \bibinfo {pages} {21} (\bibinfo {year} {2022})}\BibitemShut {NoStop}%
\bibitem [{\citenamefont {Ra}\ \emph {et~al.}(2020)\citenamefont {Ra},
  \citenamefont {Dufour}, \citenamefont {Walschaers}, \citenamefont {Jacquard},
  \citenamefont {Michel}, \citenamefont {Fabre},\ and\ \citenamefont
  {Treps}}]{Ra2020}%
  \BibitemOpen
  \bibfield  {author} {\bibinfo {author} {\bibfnamefont {Y.-S.}\ \bibnamefont
  {Ra}}, \bibinfo {author} {\bibfnamefont {A.}~\bibnamefont {Dufour}}, \bibinfo
  {author} {\bibfnamefont {M.}~\bibnamefont {Walschaers}}, \bibinfo {author}
  {\bibfnamefont {C.}~\bibnamefont {Jacquard}}, \bibinfo {author}
  {\bibfnamefont {T.}~\bibnamefont {Michel}}, \bibinfo {author} {\bibfnamefont
  {C.}~\bibnamefont {Fabre}},\ and\ \bibinfo {author} {\bibfnamefont
  {N.}~\bibnamefont {Treps}},\ }\href
  {https://doi.org/10.1038/s41567-019-0726-y} {\bibfield  {journal} {\bibinfo
  {journal} {Nature Physics}\ }\textbf {\bibinfo {volume} {16}},\ \bibinfo
  {pages} {144} (\bibinfo {year} {2020})}\BibitemShut {NoStop}%
\bibitem [{\citenamefont {Liu}\ \emph {et~al.}(2022)\citenamefont {Liu},
  \citenamefont {Han}, \citenamefont {Wang}, \citenamefont {Xiang},
  \citenamefont {Sun}, \citenamefont {Wang}, \citenamefont {Qin}, \citenamefont
  {Gong}, \citenamefont {Su},\ and\ \citenamefont
  {He}}]{PhysRevLett.128.200401}%
  \BibitemOpen
  \bibfield  {author} {\bibinfo {author} {\bibfnamefont {S.}~\bibnamefont
  {Liu}}, \bibinfo {author} {\bibfnamefont {D.}~\bibnamefont {Han}}, \bibinfo
  {author} {\bibfnamefont {N.}~\bibnamefont {Wang}}, \bibinfo {author}
  {\bibfnamefont {Y.}~\bibnamefont {Xiang}}, \bibinfo {author} {\bibfnamefont
  {F.}~\bibnamefont {Sun}}, \bibinfo {author} {\bibfnamefont {M.}~\bibnamefont
  {Wang}}, \bibinfo {author} {\bibfnamefont {Z.}~\bibnamefont {Qin}}, \bibinfo
  {author} {\bibfnamefont {Q.}~\bibnamefont {Gong}}, \bibinfo {author}
  {\bibfnamefont {X.}~\bibnamefont {Su}},\ and\ \bibinfo {author}
  {\bibfnamefont {Q.}~\bibnamefont {He}},\ }\href
  {https://doi.org/10.1103/PhysRevLett.128.200401} {\bibfield  {journal}
  {\bibinfo  {journal} {Phys. Rev. Lett.}\ }\textbf {\bibinfo {volume} {128}},\
  \bibinfo {pages} {200401} (\bibinfo {year} {2022})}\BibitemShut {NoStop}%
\bibitem [{\citenamefont {Wigner}(1932)}]{PhysRev.40.749}%
  \BibitemOpen
  \bibfield  {author} {\bibinfo {author} {\bibfnamefont {E.}~\bibnamefont
  {Wigner}},\ }\href {https://doi.org/10.1103/PhysRev.40.749} {\bibfield
  {journal} {\bibinfo  {journal} {Phys. Rev.}\ }\textbf {\bibinfo {volume}
  {40}},\ \bibinfo {pages} {749} (\bibinfo {year} {1932})}\BibitemShut
  {NoStop}%
\bibitem [{\citenamefont {Cahill}\ and\ \citenamefont
  {Glauber}(1969)}]{PhysRev.177.1882}%
  \BibitemOpen
  \bibfield  {author} {\bibinfo {author} {\bibfnamefont {K.~E.}\ \bibnamefont
  {Cahill}}\ and\ \bibinfo {author} {\bibfnamefont {R.~J.}\ \bibnamefont
  {Glauber}},\ }\href {https://doi.org/10.1103/PhysRev.177.1882} {\bibfield
  {journal} {\bibinfo  {journal} {Phys. Rev.}\ }\textbf {\bibinfo {volume}
  {177}},\ \bibinfo {pages} {1882} (\bibinfo {year} {1969})}\BibitemShut
  {NoStop}%
\bibitem [{\citenamefont {Hillery}\ \emph {et~al.}(1984)\citenamefont
  {Hillery}, \citenamefont {O'Connell}, \citenamefont {Scully},\ and\
  \citenamefont {Wigner}}]{HILLERY1984121}%
  \BibitemOpen
  \bibfield  {author} {\bibinfo {author} {\bibfnamefont {M.}~\bibnamefont
  {Hillery}}, \bibinfo {author} {\bibfnamefont {R.}~\bibnamefont {O'Connell}},
  \bibinfo {author} {\bibfnamefont {M.}~\bibnamefont {Scully}},\ and\ \bibinfo
  {author} {\bibfnamefont {E.}~\bibnamefont {Wigner}},\ }\href
  {https://doi.org/https://doi.org/10.1016/0370-1573(84)90160-1} {\bibfield
  {journal} {\bibinfo  {journal} {Physics Reports}\ }\textbf {\bibinfo {volume}
  {106}},\ \bibinfo {pages} {121} (\bibinfo {year} {1984})}\BibitemShut
  {NoStop}%
\bibitem [{\citenamefont {Holevo}(2001)}]{holevo_statistical_2001}%
  \BibitemOpen
  \bibfield  {author} {\bibinfo {author} {\bibfnamefont {A.~S.}\ \bibnamefont
  {Holevo}},\ }\href {https://doi.org/10.1007/3-540-44998-1} {\emph {\bibinfo
  {title} {Statistical {Structure} of {Quantum} {Theory}}}},\ \bibinfo {series}
  {Lecture {Notes} in {Physics} {Monographs}}, Vol.~\bibinfo {volume} {67}\
  (\bibinfo  {publisher} {Springer Berlin Heidelberg},\ \bibinfo {address}
  {Berlin, Heidelberg},\ \bibinfo {year} {2001})\BibitemShut {NoStop}%
\bibitem [{\citenamefont {Schrödinger}(1935)}]{Schrod1935}%
  \BibitemOpen
  \bibfield  {author} {\bibinfo {author} {\bibfnamefont {E.}~\bibnamefont
  {Schrödinger}},\ }\href {https://doi.org/10.1017/S0305004100013554}
  {\bibfield  {journal} {\bibinfo  {journal} {Mathematical Proceedings of the
  Cambridge Philosophical Society}\ }\textbf {\bibinfo {volume} {31}},\
  \bibinfo {pages} {555–563} (\bibinfo {year} {1935})}\BibitemShut {NoStop}%
\bibitem [{\citenamefont {Schrödinger}(1936)}]{Schrod1936}%
  \BibitemOpen
  \bibfield  {author} {\bibinfo {author} {\bibfnamefont {E.}~\bibnamefont
  {Schrödinger}},\ }\href {https://doi.org/10.1017/S0305004100019137}
  {\bibfield  {journal} {\bibinfo  {journal} {Mathematical Proceedings of the
  Cambridge Philosophical Society}\ }\textbf {\bibinfo {volume} {32}},\
  \bibinfo {pages} {446–452} (\bibinfo {year} {1936})}\BibitemShut {NoStop}%
\bibitem [{\citenamefont {Cavalcanti}\ \emph {et~al.}(2009)\citenamefont
  {Cavalcanti}, \citenamefont {Jones}, \citenamefont {Wiseman},\ and\
  \citenamefont {Reid}}]{Reid2009}%
  \BibitemOpen
  \bibfield  {author} {\bibinfo {author} {\bibfnamefont {E.~G.}\ \bibnamefont
  {Cavalcanti}}, \bibinfo {author} {\bibfnamefont {S.~J.}\ \bibnamefont
  {Jones}}, \bibinfo {author} {\bibfnamefont {H.~M.}\ \bibnamefont {Wiseman}},\
  and\ \bibinfo {author} {\bibfnamefont {M.~D.}\ \bibnamefont {Reid}},\ }\href
  {https://doi.org/10.1103/PhysRevA.80.032112} {\bibfield  {journal} {\bibinfo
  {journal} {Phys. Rev. A}\ }\textbf {\bibinfo {volume} {80}},\ \bibinfo
  {pages} {032112} (\bibinfo {year} {2009})}\BibitemShut {NoStop}%
\bibitem [{\citenamefont {Reid}\ \emph {et~al.}(2009)\citenamefont {Reid},
  \citenamefont {Drummond}, \citenamefont {Bowen}, \citenamefont {Cavalcanti},
  \citenamefont {Lam}, \citenamefont {Bachor}, \citenamefont {Andersen},\ and\
  \citenamefont {Leuchs}}]{Reid2009Colloquium}%
  \BibitemOpen
  \bibfield  {author} {\bibinfo {author} {\bibfnamefont {M.~D.}\ \bibnamefont
  {Reid}}, \bibinfo {author} {\bibfnamefont {P.~D.}\ \bibnamefont {Drummond}},
  \bibinfo {author} {\bibfnamefont {W.~P.}\ \bibnamefont {Bowen}}, \bibinfo
  {author} {\bibfnamefont {E.~G.}\ \bibnamefont {Cavalcanti}}, \bibinfo
  {author} {\bibfnamefont {P.~K.}\ \bibnamefont {Lam}}, \bibinfo {author}
  {\bibfnamefont {H.~A.}\ \bibnamefont {Bachor}}, \bibinfo {author}
  {\bibfnamefont {U.~L.}\ \bibnamefont {Andersen}},\ and\ \bibinfo {author}
  {\bibfnamefont {G.}~\bibnamefont {Leuchs}},\ }\href
  {https://doi.org/10.1103/RevModPhys.81.1727} {\bibfield  {journal} {\bibinfo
  {journal} {Rev. Mod. Phys.}\ }\textbf {\bibinfo {volume} {81}},\ \bibinfo
  {pages} {1727} (\bibinfo {year} {2009})}\BibitemShut {NoStop}%
\bibitem [{\citenamefont {Wiseman}\ \emph {et~al.}(2007)\citenamefont
  {Wiseman}, \citenamefont {Jones},\ and\ \citenamefont
  {Doherty}}]{Wiseman2007}%
  \BibitemOpen
  \bibfield  {author} {\bibinfo {author} {\bibfnamefont {H.~M.}\ \bibnamefont
  {Wiseman}}, \bibinfo {author} {\bibfnamefont {S.~J.}\ \bibnamefont {Jones}},\
  and\ \bibinfo {author} {\bibfnamefont {A.~C.}\ \bibnamefont {Doherty}},\
  }\href {https://doi.org/10.1103/PhysRevLett.98.140402} {\bibfield  {journal}
  {\bibinfo  {journal} {Phys. Rev. Lett.}\ }\textbf {\bibinfo {volume} {98}},\
  \bibinfo {pages} {140402} (\bibinfo {year} {2007})}\BibitemShut {NoStop}%
\bibitem [{\citenamefont {Cavalcanti}\ and\ \citenamefont
  {Skrzypczyk}(2016)}]{Cavalcanti2016}%
  \BibitemOpen
  \bibfield  {author} {\bibinfo {author} {\bibfnamefont {D.}~\bibnamefont
  {Cavalcanti}}\ and\ \bibinfo {author} {\bibfnamefont {P.}~\bibnamefont
  {Skrzypczyk}},\ }\href {https://doi.org/10.1088/1361-6633/80/2/024001}
  {\bibfield  {journal} {\bibinfo  {journal} {Reports on Progress in Physics}\
  }\textbf {\bibinfo {volume} {80}},\ \bibinfo {pages} {024001} (\bibinfo
  {year} {2016})}\BibitemShut {NoStop}%
\bibitem [{\citenamefont {Uola}\ \emph {et~al.}(2020)\citenamefont {Uola},
  \citenamefont {Costa}, \citenamefont {Nguyen},\ and\ \citenamefont
  {G\"uhne}}]{Uola2020}%
  \BibitemOpen
  \bibfield  {author} {\bibinfo {author} {\bibfnamefont {R.}~\bibnamefont
  {Uola}}, \bibinfo {author} {\bibfnamefont {A.~C.~S.}\ \bibnamefont {Costa}},
  \bibinfo {author} {\bibfnamefont {H.~C.}\ \bibnamefont {Nguyen}},\ and\
  \bibinfo {author} {\bibfnamefont {O.}~\bibnamefont {G\"uhne}},\ }\href
  {https://doi.org/10.1103/RevModPhys.92.015001} {\bibfield  {journal}
  {\bibinfo  {journal} {Rev. Mod. Phys.}\ }\textbf {\bibinfo {volume} {92}},\
  \bibinfo {pages} {015001} (\bibinfo {year} {2020})}\BibitemShut {NoStop}%
\bibitem [{\citenamefont {Schneeloch}\ \emph {et~al.}(2013)\citenamefont
  {Schneeloch}, \citenamefont {Broadbent}, \citenamefont {Walborn},
  \citenamefont {Cavalcanti},\ and\ \citenamefont
  {Howell}}]{PhysRevA.87.062103}%
  \BibitemOpen
  \bibfield  {author} {\bibinfo {author} {\bibfnamefont {J.}~\bibnamefont
  {Schneeloch}}, \bibinfo {author} {\bibfnamefont {C.~J.}\ \bibnamefont
  {Broadbent}}, \bibinfo {author} {\bibfnamefont {S.~P.}\ \bibnamefont
  {Walborn}}, \bibinfo {author} {\bibfnamefont {E.~G.}\ \bibnamefont
  {Cavalcanti}},\ and\ \bibinfo {author} {\bibfnamefont {J.~C.}\ \bibnamefont
  {Howell}},\ }\href {https://doi.org/10.1103/PhysRevA.87.062103} {\bibfield
  {journal} {\bibinfo  {journal} {Phys. Rev. A}\ }\textbf {\bibinfo {volume}
  {87}},\ \bibinfo {pages} {062103} (\bibinfo {year} {2013})}\BibitemShut
  {NoStop}%
\bibitem [{\citenamefont {Yadin}\ \emph {et~al.}(2021)\citenamefont {Yadin},
  \citenamefont {Fadel},\ and\ \citenamefont {Gessner}}]{YFG21}%
  \BibitemOpen
  \bibfield  {author} {\bibinfo {author} {\bibfnamefont {B.}~\bibnamefont
  {Yadin}}, \bibinfo {author} {\bibfnamefont {M.}~\bibnamefont {Fadel}},\ and\
  \bibinfo {author} {\bibfnamefont {M.}~\bibnamefont {Gessner}},\ }\href
  {https://doi.org/10.1038/s41467-021-22353-3} {\bibfield  {journal} {\bibinfo
  {journal} {Nature Communications}\ }\textbf {\bibinfo {volume} {12}},\
  \bibinfo {pages} {2410} (\bibinfo {year} {2021})}\BibitemShut {NoStop}%
\bibitem [{\citenamefont {Lopetegui}\ \emph {et~al.}(2022)\citenamefont
  {Lopetegui}, \citenamefont {Gessner}, \citenamefont {Fadel}, \citenamefont
  {Treps},\ and\ \citenamefont {Walschaers}}]{PRXQuantum.3.030347}%
  \BibitemOpen
  \bibfield  {author} {\bibinfo {author} {\bibfnamefont {C.~E.}\ \bibnamefont
  {Lopetegui}}, \bibinfo {author} {\bibfnamefont {M.}~\bibnamefont {Gessner}},
  \bibinfo {author} {\bibfnamefont {M.}~\bibnamefont {Fadel}}, \bibinfo
  {author} {\bibfnamefont {N.}~\bibnamefont {Treps}},\ and\ \bibinfo {author}
  {\bibfnamefont {M.}~\bibnamefont {Walschaers}},\ }\href
  {https://doi.org/10.1103/PRXQuantum.3.030347} {\bibfield  {journal} {\bibinfo
   {journal} {PRX Quantum}\ }\textbf {\bibinfo {volume} {3}},\ \bibinfo {pages}
  {030347} (\bibinfo {year} {2022})}\BibitemShut {NoStop}%
\bibitem [{\citenamefont {Cavaill\`es}\ \emph {et~al.}(2018)\citenamefont
  {Cavaill\`es}, \citenamefont {Le~Jeannic}, \citenamefont {Raskop},
  \citenamefont {Guccione}, \citenamefont {Markham}, \citenamefont {Diamanti},
  \citenamefont {Shaw}, \citenamefont {Verma}, \citenamefont {Nam},\ and\
  \citenamefont {Laurat}}]{Laurat2018}%
  \BibitemOpen
  \bibfield  {author} {\bibinfo {author} {\bibfnamefont {A.}~\bibnamefont
  {Cavaill\`es}}, \bibinfo {author} {\bibfnamefont {H.}~\bibnamefont
  {Le~Jeannic}}, \bibinfo {author} {\bibfnamefont {J.}~\bibnamefont {Raskop}},
  \bibinfo {author} {\bibfnamefont {G.}~\bibnamefont {Guccione}}, \bibinfo
  {author} {\bibfnamefont {D.}~\bibnamefont {Markham}}, \bibinfo {author}
  {\bibfnamefont {E.}~\bibnamefont {Diamanti}}, \bibinfo {author}
  {\bibfnamefont {M.~D.}\ \bibnamefont {Shaw}}, \bibinfo {author}
  {\bibfnamefont {V.~B.}\ \bibnamefont {Verma}}, \bibinfo {author}
  {\bibfnamefont {S.~W.}\ \bibnamefont {Nam}},\ and\ \bibinfo {author}
  {\bibfnamefont {J.}~\bibnamefont {Laurat}},\ }\href
  {https://doi.org/10.1103/PhysRevLett.121.170403} {\bibfield  {journal}
  {\bibinfo  {journal} {Phys. Rev. Lett.}\ }\textbf {\bibinfo {volume} {121}},\
  \bibinfo {pages} {170403} (\bibinfo {year} {2018})}\BibitemShut {NoStop}%
\bibitem [{\citenamefont {Lami}\ \emph {et~al.}(2016)\citenamefont {Lami},
  \citenamefont {Hirche}, \citenamefont {Adesso},\ and\ \citenamefont
  {Winter}}]{PhysRevLett.117.220502}%
  \BibitemOpen
  \bibfield  {author} {\bibinfo {author} {\bibfnamefont {L.}~\bibnamefont
  {Lami}}, \bibinfo {author} {\bibfnamefont {C.}~\bibnamefont {Hirche}},
  \bibinfo {author} {\bibfnamefont {G.}~\bibnamefont {Adesso}},\ and\ \bibinfo
  {author} {\bibfnamefont {A.}~\bibnamefont {Winter}},\ }\href
  {https://doi.org/10.1103/PhysRevLett.117.220502} {\bibfield  {journal}
  {\bibinfo  {journal} {Phys. Rev. Lett.}\ }\textbf {\bibinfo {volume} {117}},\
  \bibinfo {pages} {220502} (\bibinfo {year} {2016})}\BibitemShut {NoStop}%
\bibitem [{\citenamefont {Lami}\ \emph {et~al.}(2018)\citenamefont {Lami},
  \citenamefont {Serafini},\ and\ \citenamefont {Adesso}}]{Lami_2018}%
  \BibitemOpen
  \bibfield  {author} {\bibinfo {author} {\bibfnamefont {L.}~\bibnamefont
  {Lami}}, \bibinfo {author} {\bibfnamefont {A.}~\bibnamefont {Serafini}},\
  and\ \bibinfo {author} {\bibfnamefont {G.}~\bibnamefont {Adesso}},\ }\href
  {https://doi.org/10.1088/1367-2630/aaa654} {\bibfield  {journal} {\bibinfo
  {journal} {New Journal of Physics}\ }\textbf {\bibinfo {volume} {20}},\
  \bibinfo {pages} {023030} (\bibinfo {year} {2018})}\BibitemShut {NoStop}%
\bibitem [{\citenamefont {Kogias}\ \emph
  {et~al.}(2015{\natexlab{a}})\citenamefont {Kogias}, \citenamefont {Lee},
  \citenamefont {Ragy},\ and\ \citenamefont {Adesso}}]{PhysRevLett.114.060403}%
  \BibitemOpen
  \bibfield  {author} {\bibinfo {author} {\bibfnamefont {I.}~\bibnamefont
  {Kogias}}, \bibinfo {author} {\bibfnamefont {A.~R.}\ \bibnamefont {Lee}},
  \bibinfo {author} {\bibfnamefont {S.}~\bibnamefont {Ragy}},\ and\ \bibinfo
  {author} {\bibfnamefont {G.}~\bibnamefont {Adesso}},\ }\href
  {https://doi.org/10.1103/PhysRevLett.114.060403} {\bibfield  {journal}
  {\bibinfo  {journal} {Phys. Rev. Lett.}\ }\textbf {\bibinfo {volume} {114}},\
  \bibinfo {pages} {060403} (\bibinfo {year} {2015}{\natexlab{a}})}\BibitemShut
  {NoStop}%
\bibitem [{\citenamefont {Kogias}\ \emph
  {et~al.}(2015{\natexlab{b}})\citenamefont {Kogias}, \citenamefont
  {Skrzypczyk}, \citenamefont {Cavalcanti}, \citenamefont {Ac\'{\i}n},\ and\
  \citenamefont {Adesso}}]{PhysRevLett.115.210401}%
  \BibitemOpen
  \bibfield  {author} {\bibinfo {author} {\bibfnamefont {I.}~\bibnamefont
  {Kogias}}, \bibinfo {author} {\bibfnamefont {P.}~\bibnamefont {Skrzypczyk}},
  \bibinfo {author} {\bibfnamefont {D.}~\bibnamefont {Cavalcanti}}, \bibinfo
  {author} {\bibfnamefont {A.}~\bibnamefont {Ac\'{\i}n}},\ and\ \bibinfo
  {author} {\bibfnamefont {G.}~\bibnamefont {Adesso}},\ }\href
  {https://doi.org/10.1103/PhysRevLett.115.210401} {\bibfield  {journal}
  {\bibinfo  {journal} {Phys. Rev. Lett.}\ }\textbf {\bibinfo {volume} {115}},\
  \bibinfo {pages} {210401} (\bibinfo {year} {2015}{\natexlab{b}})}\BibitemShut
  {NoStop}%
\bibitem [{\citenamefont {Einstein}\ \emph {et~al.}(1935)\citenamefont
  {Einstein}, \citenamefont {Podolsky},\ and\ \citenamefont {Rosen}}]{EPR1935}%
  \BibitemOpen
  \bibfield  {author} {\bibinfo {author} {\bibfnamefont {A.}~\bibnamefont
  {Einstein}}, \bibinfo {author} {\bibfnamefont {B.}~\bibnamefont {Podolsky}},\
  and\ \bibinfo {author} {\bibfnamefont {N.}~\bibnamefont {Rosen}},\ }\href
  {https://doi.org/10.1103/PhysRev.47.777} {\bibfield  {journal} {\bibinfo
  {journal} {Phys. Rev.}\ }\textbf {\bibinfo {volume} {47}},\ \bibinfo {pages}
  {777} (\bibinfo {year} {1935})}\BibitemShut {NoStop}%
\bibitem [{\citenamefont {Reid}(1989)}]{PhysRevA.40.913}%
  \BibitemOpen
  \bibfield  {author} {\bibinfo {author} {\bibfnamefont {M.~D.}\ \bibnamefont
  {Reid}},\ }\href {https://doi.org/10.1103/PhysRevA.40.913} {\bibfield
  {journal} {\bibinfo  {journal} {Phys. Rev. A}\ }\textbf {\bibinfo {volume}
  {40}},\ \bibinfo {pages} {913} (\bibinfo {year} {1989})}\BibitemShut
  {NoStop}%
\bibitem [{\citenamefont {Chabaud}\ \emph {et~al.}(2020)\citenamefont
  {Chabaud}, \citenamefont {Markham},\ and\ \citenamefont
  {Grosshans}}]{PhysRevLett.124.063605}%
  \BibitemOpen
  \bibfield  {author} {\bibinfo {author} {\bibfnamefont {U.}~\bibnamefont
  {Chabaud}}, \bibinfo {author} {\bibfnamefont {D.}~\bibnamefont {Markham}},\
  and\ \bibinfo {author} {\bibfnamefont {F.}~\bibnamefont {Grosshans}},\ }\href
  {https://doi.org/10.1103/PhysRevLett.124.063605} {\bibfield  {journal}
  {\bibinfo  {journal} {Phys. Rev. Lett.}\ }\textbf {\bibinfo {volume} {124}},\
  \bibinfo {pages} {063605} (\bibinfo {year} {2020})}\BibitemShut {NoStop}%
\bibitem [{\citenamefont {Filip}\ and\ \citenamefont {Mi\ifmmode~\check{s}\else
  \v{s}\fi{}ta}(2011)}]{PhysRevLett.106.200401}%
  \BibitemOpen
  \bibfield  {author} {\bibinfo {author} {\bibfnamefont {R.}~\bibnamefont
  {Filip}}\ and\ \bibinfo {author} {\bibfnamefont {L.}~\bibnamefont
  {Mi\ifmmode~\check{s}\else \v{s}\fi{}ta}},\ }\href
  {https://doi.org/10.1103/PhysRevLett.106.200401} {\bibfield  {journal}
  {\bibinfo  {journal} {Phys. Rev. Lett.}\ }\textbf {\bibinfo {volume} {106}},\
  \bibinfo {pages} {200401} (\bibinfo {year} {2011})}\BibitemShut {NoStop}%
\bibitem [{\citenamefont {Ferrie}(2011)}]{Ferrie_2011}%
  \BibitemOpen
  \bibfield  {author} {\bibinfo {author} {\bibfnamefont {C.}~\bibnamefont
  {Ferrie}},\ }\href {https://doi.org/10.1088/0034-4885/74/11/116001}
  {\bibfield  {journal} {\bibinfo  {journal} {Reports on Progress in Physics}\
  }\textbf {\bibinfo {volume} {74}},\ \bibinfo {pages} {116001} (\bibinfo
  {year} {2011})}\BibitemShut {NoStop}%
\bibitem [{\citenamefont {Booth}\ \emph {et~al.}(2022)\citenamefont {Booth},
  \citenamefont {Chabaud},\ and\ \citenamefont {Emeriau}}]{Contextuality1}%
  \BibitemOpen
  \bibfield  {author} {\bibinfo {author} {\bibfnamefont {R.~I.}\ \bibnamefont
  {Booth}}, \bibinfo {author} {\bibfnamefont {U.}~\bibnamefont {Chabaud}},\
  and\ \bibinfo {author} {\bibfnamefont {P.-E.}\ \bibnamefont {Emeriau}},\
  }\href {https://doi.org/10.1103/PhysRevLett.129.230401} {\bibfield  {journal}
  {\bibinfo  {journal} {Phys. Rev. Lett.}\ }\textbf {\bibinfo {volume} {129}},\
  \bibinfo {pages} {230401} (\bibinfo {year} {2022})}\BibitemShut {NoStop}%
\bibitem [{\citenamefont {Haferkamp}\ and\ \citenamefont
  {Bermejo-Vega}(2021)}]{Contextuality2}%
  \BibitemOpen
  \bibfield  {author} {\bibinfo {author} {\bibfnamefont {J.}~\bibnamefont
  {Haferkamp}}\ and\ \bibinfo {author} {\bibfnamefont {J.}~\bibnamefont
  {Bermejo-Vega}},\ }\href {https://doi.org/10.48550/ARXIV.2112.14788}
  {\bibinfo {title} {Equivalence of contextuality and wigner function
  negativity in continuous-variable quantum optics}} (\bibinfo {year}
  {2021})\BibitemShut {NoStop}%
\end{thebibliography}%

\end{document}